# The Dual Nature of Body-Axis Formation in *Hydra* Regeneration: Polarity-Morphology Concurrency

Oded Agam[1] and Erez Braun[2]

[1]The Racah Institute of Physics, Edmond J. Safra Campus, The Hebrew University of Jerusalem, Jerusalem 9190401, Israel.

[2] Department of Physics and Network Biology Research Laboratories, Technion-Israel Institute of Technology, Haifa 32000, Israel.

## ABSTRACT

The formation of a body-axis is central to animal development and involves both polarity and morphology. While polarity is traditionally associated with biochemical patterning, the morphological aspect of axis formation remains elusive. In regenerating *Hydra* tissues, we find that morphological evolution in all tissue samples depends on inherited positional information from the donor's axis, and a foot precursor emerges early in the process. From the onset of regeneration, the $Ca^{2+}$ excitations that drive actomyosin forces for tissue reshaping follow a gradient aligned with the head-foot polarity direction. We conclude that polarity and morphological axis progression occur concurrently through interlinked processes, and that the foot plays a dominant role in this process, a role usually attributed to the head organizer. A simple toy model accounts for the observed regeneration dynamics and illustrates the mechanochemical integration of polarity and morphogenesis. We expect the insights from *Hydra* to be relevant to broader developmental systems.

Morphogenesis in animal development arises from a complex interplay of biochemical, mechanical, and electrical processes that act across multiple scales, from subcellular dynamics to the whole organism[1-11]. Identifying the organizational principles that integrate and coordinate these processes for robust developmental outcomes remains challenging. The emergence of a body axis, which provides an organizational skeleton for development, is a prominent manifestation of symmetry-breaking events[12, 13]. Yet, the internal anatomical axis may not necessarily mirror external morphological symmetry. In bilaterians, for example, the left-right (LR) axis specifies internal asymmetries despite bilateral external symmetry[14, 15]. Thus, axis formation encompasses two processes: polarity, manifested as symmetry-breaking gradients driven by biochemical signals that provide positional cues, and morphological evolution, which shapes the body form. Traditional studies emphasize polarity, but mechanical processes, which are vital for morphological-axis formation, remain less understood[2, 16, 17]. These two aspects of the body-axis, polarity and morphology, must be interlinked to ensure coordinated animal development[18].

Here, we study this coordination using whole-body *Hydra* regeneration from tissue fragments. *Hydra*, a freshwater cnidarian with a single well-defined axis, exhibits robust morphogenesis and remarkable regenerative capacity[4, 9, 19-23]. When a flat tissue fragment of several hundred cells is excised from a mature *Hydra*, it first folds into a closed hollow sphere before regenerating into a complete animal[4]. Regeneration primarily involves changes in cell shape, tissue reorganization, and differentiation, whereas cell division is not strictly required[22, 24, 25]. The head organizer in *Hydra* is considered to play an important role in establishing and maintaining polarity during regeneration, as well as during the continuous tissue renewal process of the mature animal that never ceases [19, 26-29]. When a mature *Hydra* is bisected or a tissue fragment is excised, the resulting polarity vector (deduced indirectly by transplantation and grafting experiments) is inherited from the parent animal [30-34]. Moreover, the regeneration dynamics of excised tissue segments as well as their induction effects when transplanted, retain a “positional memory” of their original location along the parent *Hydra*’s body axis relative to the head organizer [26, 35]. These observations highlight the head organizer’s role in axis polarity, with the *Wnt*/*β-catenin* signaling as a key player and *Wnt3* serving as the main inductive signal[27, 28, 33, 34, 36-39].

The primary morphological transition in *Hydra* regeneration, from a spherical shape to a tube-like form, marks the emergence of a morphological axis[8]. The *Hydra*’s body behaves like a soft muscle due to supracellular actin fibers in the epithelium[23, 40-42]. On the timescale of the morphological transition, shape changes are mainly driven by active contractile forces generated by the actomyosin fibers[4] and by hydrostatic pressure within the tissue’s cavity, modulated by osmotic gradients[34, 43, 44]. Although tissue fragments inherit a partial alignment of actin fibers from the parent polyp [4, 23], the resulting distribution of mechanical forces depends on the local concentration of active myosin and on $Ca^{2+}$ excitations. Indeed, our previous work shows that $Ca^{2+}$ activity, coupled to tissue curvature, is more influential in determining morphology than mere actin-fiber alignment[8, 45]. Further experimental studies demonstrate that the actin fiber organization can adapt to biochemical or mechanical constraints[44, 46], suggesting that these fibers might actually follow, rather than define, the emerging axis.

The main goal of this study is to shed light on morphological-axis formation and to explore whether morphology and polarity are two facets of a single process underlying body-axis establishment, or two interdependent processes that proceed in parallel. Towards this goal, we examine the trajectories of various tissue samples in morphological space, each with distinct initial conditions defined by the tissue size and its excision site along the parent *Hydra*’s body-axis. Our findings indicate that polarity establishment and morphological evolution arise from separate, yet interdependent processes that progress concurrently. In addition, $Ca^{2+}$ excitations, which are critical regulators of morphological changes, show a clear continuous gradient profile which is aligned with the polarity vector, from the onset of a folded excised tissue fragment. Thus, the $Ca^{2+}$ field provides a dynamic link between polarity and morphological-axis formation, orchestrating shape changes along the polarity axis to ensure robust regeneration. We propose a framework, illustrated by a simple toy model, in which inherited positional factors from the parent *Hydra*, encoding the tissue's original distance from the head or foot, dynamically regulate both the mechanical properties of the tissue and its local $Ca^{2+}$ activity. The interplay between these factors establishes a stable calcium gradient aligned with the original polarity axis and dictates the morphological trajectory leading to the regenerated *Hydra* body form.

**Initial Conditions Shape Morphological Dynamics**

We first analyze the trajectories of regenerating tissue fragments in morphological space, each defined by different initial conditions (Fig. 1a). For a concise overview of how tissues evolve morphologically under these varying conditions, we apply a t-distributed Stochastic Neighbor Embedding (t-SNE) dimensionality reduction analysis (Fig. 1c)[47]. Each sample's trajectory in morphological space is characterized by the transition duration from a spherical to a cylindrical-like shape, and by the magnitude of shape fluctuations, quantified through the leading harmonic moments of the projected tissue image[48, 49] (see Eq. (1) below, and Supplementary Note 1). The clustering of tissue samples according to their initial conditions indicates that both tissue size and positional information - i.e., memory of the excision site along the parent *Hydra*'s axis - play a dominant role in directing the system's morphological evolution. For details on the dimensionality reduction procedure and an alternative approach (UMAP[50]) yielding similar results, see Supplementary Note 2 and Fig. S1.

**Quantifying Tissue Shape Dynamics**

We next characterize the tissue shape changes over time using a shape parameter, $\Lambda = 1 - 4\pi A / P^2$, where $A$ is the area and $P$ is the perimeter of the projected tissue shape[8, 45]. This parameter is zero for a spherical tissue and becomes nonzero whenever the projected image deviates from a perfect circle. During the initial stages of regeneration, the *Hydra*'s body axis typically remains parallel to the imaging plane, deviating out of plane only after the foot is fully developed. Thus, $\Lambda$ provides a reliable measure of tissue elongation, and its persistent deviation from zero marks the establishment of the morphological axis. However, $\Lambda$ characterizes only deviations of the projected image from a circular shape, which is insufficient for capturing the full complexity of the tissue's morphology. To address this limitation, we introduce a second measure: the external harmonic moments of the projected *Hydra* image, a method extensively used to describe two-dimensional patterns, defined as[48, 49]:

$$t_n = \frac{1}{\pi n} \iint_D \frac{dxdy}{\left(x+iy\right)^n}, \quad n = 1, 2, .... \tag{1}$$

where $x$ and $y$ denote the coordinates in an arbitrary coordinate system defined on the image projection plane, and the integration domain, $D$, corresponds to the exterior of the projected *Hydra* image. It can be demonstrated that the infinite set of harmonic moments, along with the projected image area, uniquely determine the image shape[48, 49]. The first harmonic moment, $t_1$, can always be set to zero by appropriately

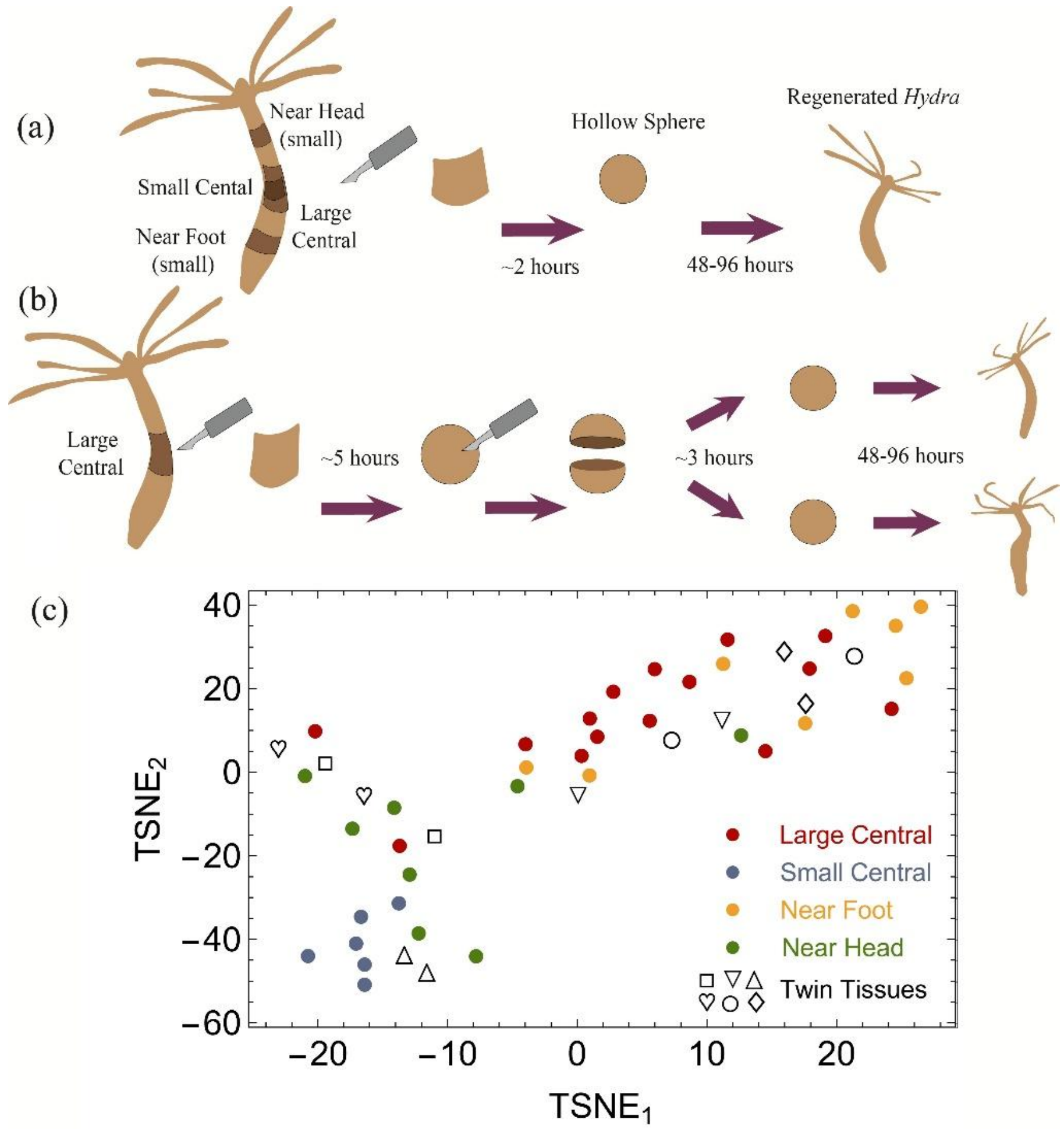

**Fig. 1. Experimental protocols and corresponding morphological evolution classification.** (a) The experimental procedures used to establish different initial conditions for regenerating tissue samples, accounting for both positional and size dependencies. Tissue fragments are excised from different locations along a mature *Hydra*. The flat tissue fragments typically fold into closed hollow spheres within approximately 2-3 hours after excision, and full regeneration occurs within 48–96 hours. After folding, each tissue fragment is transferred to the experimental setup under a time-lapse fluorescence microscope[8-10, 45]. All experiments use the same *Hydra* medium[4], capturing both bright-field (BF) and fluorescence images at one-minute intervals throughout the regeneration process (Methods). (b) Twin tissue experiments: A large tissue fragment excised from the central-axis region of a mature *Hydra*, is allowed to fold into a closed hollow sphere for ~5 hours, after which it is cut into two smaller fragments that re-fold and proceed independently toward regeneration. (c) t-SNE[47] analysis (see main text) of regeneration trajectories in morphological space for 49 tissue samples, each marked by a dot, revealing clear clustering of tissues based on their initial conditions. Samples: (red) Large tissue samples originated from the central section along the axis of a mature *Hydra*; (blue) small tissue samples excised from the center; (yellow) small tissue samples excised from regions close to the foot; (green) small tissue samples excised from regions close to the head. Black symbols denote twin tissue pairs, with matching symbols indicating samples derived from the same parent tissue fragment. Note that a twin pair is generally positioned close to each other in the reduced space, manifesting their shared characteristics.

choosing the origin of the coordinate system, a convention we adopt here. All $t_n$ values vanish for a circular domain centered at the origin. For small deviations from a circle, these moments effectively represent the

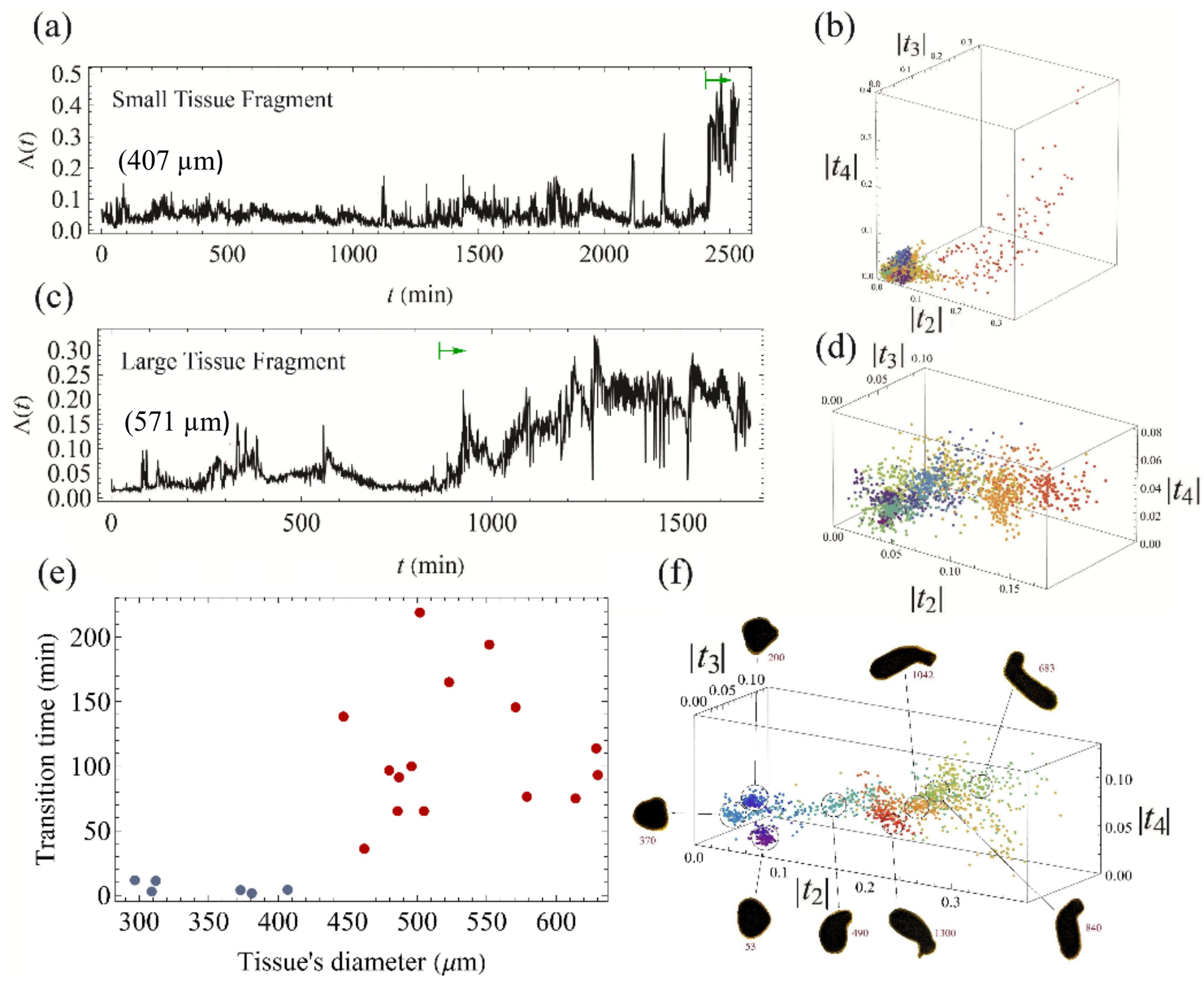

**Fig. 2. Morphological evolution of small versus large tissue fragments.** (a) Shape parameter evolution of a small tissue fragment excised from the mid-axis of the parent *Hydra*, showing a rapid morphological transition (~5 minutes) following an extended period of a nearly spherical shape. (b) Regeneration trajectory of the same small fragment in the reduced morphological space, defined by the absolute values of the leading harmonic moments (see main text), with a rainbow color code representing time (purple for early times, red for later stages). (c) Shape parameter evolution of a large tissue fragment excised from the mid-axis of the parent *Hydra*, exhibiting a slow, gradual transition with significant shape fluctuations. The green arrows in panels a and c indicate the onset of the tissue's morphological change, with time in these traces measured from the start of observations, 2–3 hours after excision. (d) Regeneration trajectory of the large fragment in the reduced morphological space. (e) The transition duration of the morphological change as a function of the projected tissue diameter (measured when the fragment is approximately spherical). Blue and red mark small and large central fragments, respectively. (f) A regeneration trajectory of a large tissue fragment in the reduced morphological space, highlighting prolonged residence in quasi-metastable states. Representative tissue images illustrate corresponding morphological states along the trajectory (numbers indicate the respective time of the image).

Fourier transform of the polar representation $r(\theta)$ of the projected image contour as a function of the angle $\theta$ (Supplementary Note 2). Thus, focusing on the first few moments provides a simplified, coarse-grained description within a reduced-dimensional morphological space.

### The Influence of Tissue Size on Morphological Evolution

Fig. 2 illustrates the regeneration dynamics of small and large tissue fragments excised from the central region of the *Hydra*'s axis, using the measures described above. Fig. 2e displays a distinct shift in the morphological transition duration for tissue samples exceeding a diameter of ~430 μm. Below this scale, the morphological transition from a nearly spherical shape to a tubular structure occurs rapidly, typically within a few minutes (Fig. 2a). In contrast, large tissue samples (diameter > 450 μm) exhibit a dramatically different behavior, characterized by three key features: (i) a gradual and prolonged morphological transition (Fig. 2c), (ii) broader and more diverse morphological fluctuations (compare Fig. 2b and 2d), and (iii) a regeneration trajectory that traverses a wider range of morphologies, absent in smaller samples (Fig. 2f). These observations were consistently reproduced across 6 independent small samples[8] and 15 large samples, collected from 5 separate experiments (see Supplementary Figures S2 & S3).

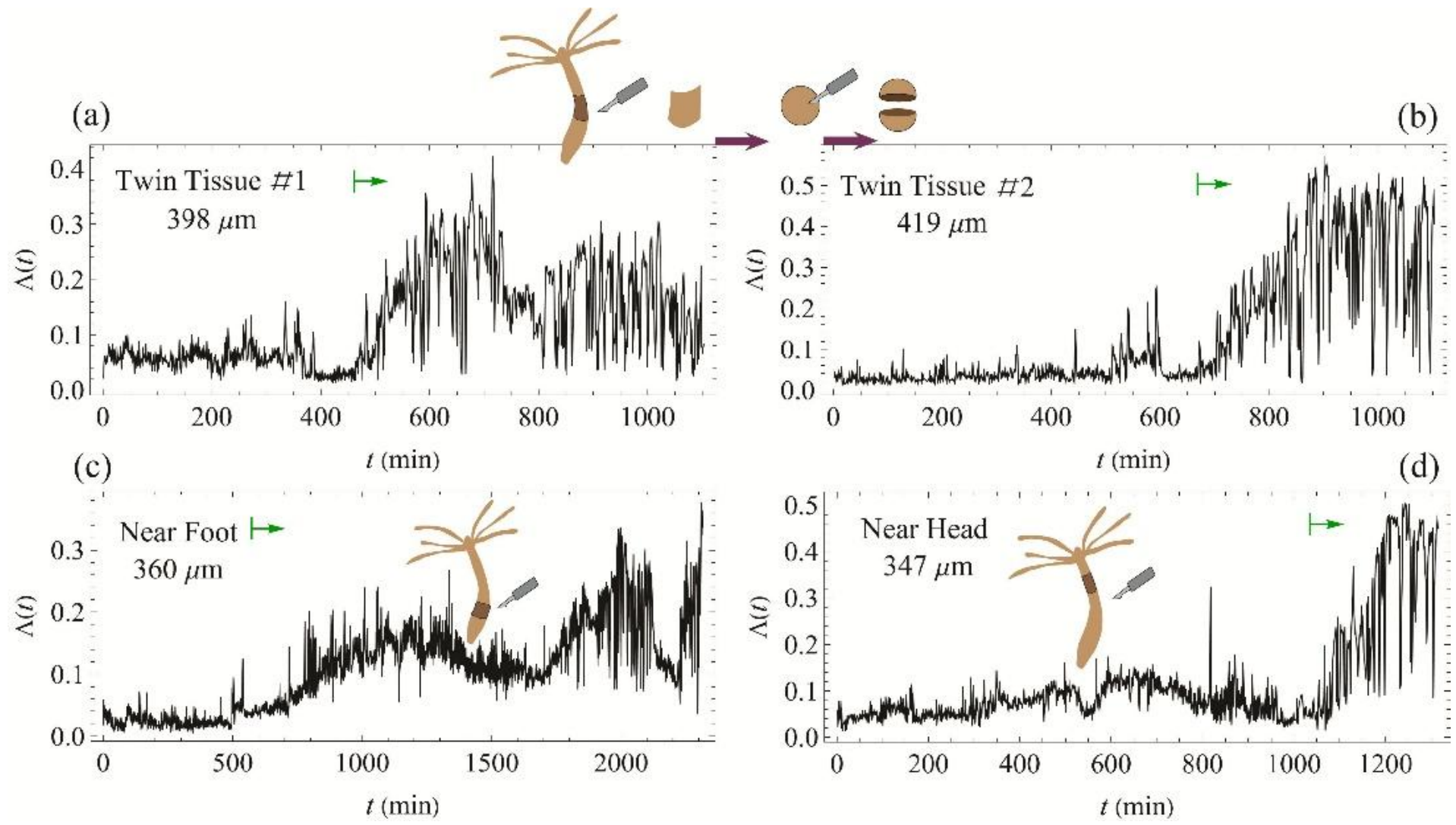

**Fig. 3. Impact of positional information on morphological evolution.** (a) & (b) Shape parameters over time for a pair of small twin tissue samples derived from the same large tissue fragment, exhibiting gradual morphological transitions accompanied by strong fluctuations. (c) & (d) Shape parameter evolution for small tissue samples excised near the foot and near the head respectively, demonstrating distinct behaviors compared to small fragments from the central body region, as shown in Fig. 2a. All tissue samples are below the critical size threshold identified in Fig. 2e. Tissue size is measured after folding into hollow spheres and reflects a balance between internal hydrostatic pressure and actomyosin forces; thus, twin tissues are typically larger than half the size of the original parent fragment. Time zero corresponds to the start of measurements, 2–3 hours after cutting (either from the large, folded tissue for twins, or directly from the parent *Hydra* near the head or foot). Green arrows indicate the onset of the tissue's morphological change.

**Positional Dependence of Morphogenesis**

To assess whether the tissue size is the primary factor influencing morphological dynamics, we conduct a "twin-tissue experiment", see Fig. 1b. A large tissue fragment (>450 μm) is excised from the central region of a mature *Hydra* and allowed to fold into a hollow closed spherical shape over approximately 5 hours. The folded tissue is then cut into two sub-fragments, each smaller than the critical-size threshold identified in Fig. 2e. These smaller sub-fragments are subsequently permitted to re-fold into closed hollow spheres for roughly 3 hours, after which their regeneration trajectories are recorded.

Figs. 3a and 3b show the shape parameter as a function of time for a representative pair of twin tissue samples. The results are striking. Although both samples are below the critical size threshold, their behavior differs significantly from that of small tissue fragments excised directly from the *Hydra*'s mid-axis. Similar patterns were observed across six pairs of twin tissue samples (see Supplementary Fig. S4). These results indicate that the tissue's size is not the primary factor affecting the regeneration dynamics. Instead, the dynamics are governed by factors inherited from the parent *Hydra* or potentially developed during the five

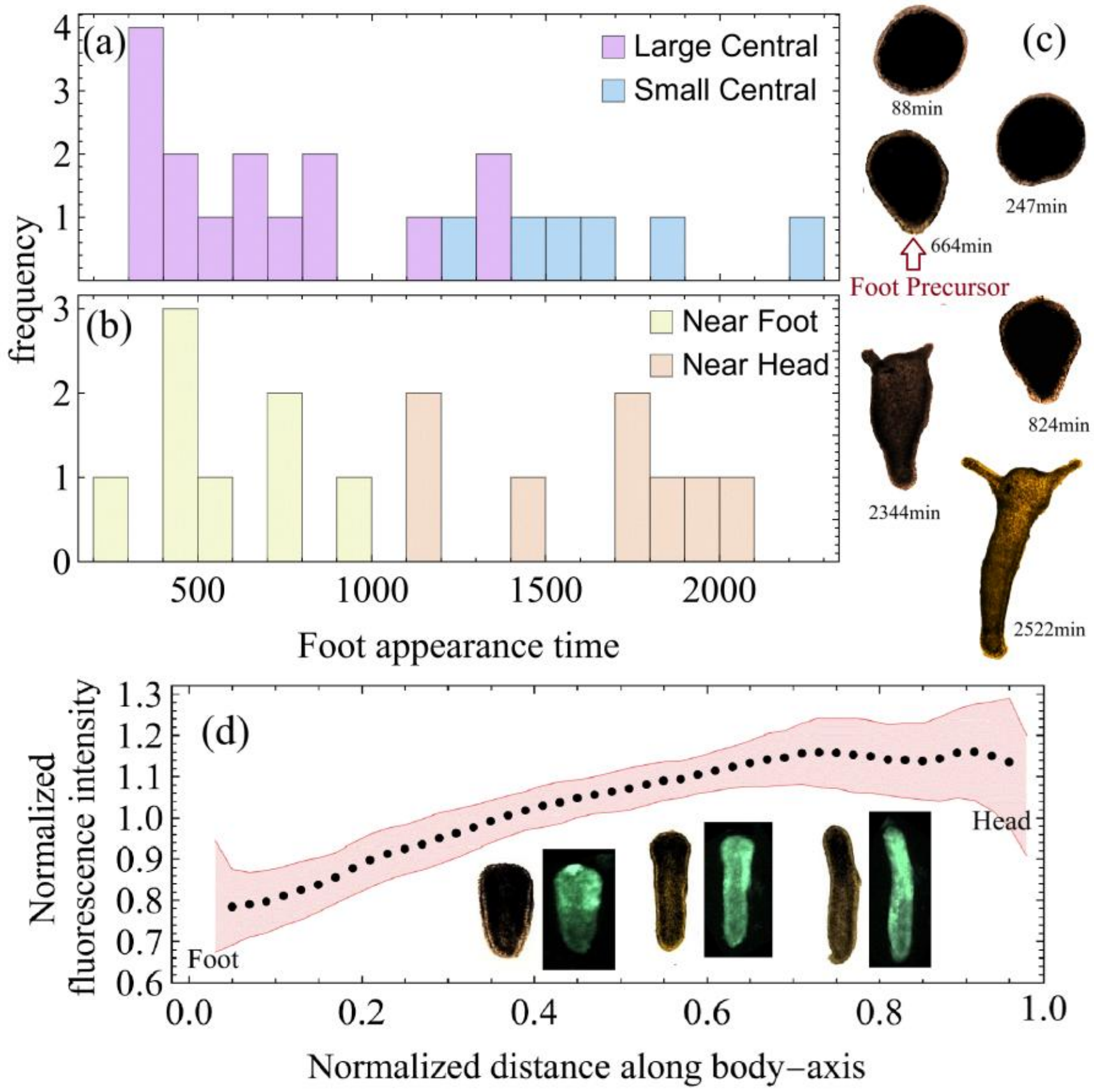

**Fig. 4. Indicators of polarity establishment.** (a) & (b) Histograms of the approximate appearance time of the foot precursor for large and small mid-axis tissues, as well as small tissues near the head and foot, respectively (time is measured from the tissue's cutting moment). Due to the limitations in analyzing projected tissue images, the observed time represents an upper limit for the first appearance of the foot precursor. (c) Bright-field (BF) images of a regenerating tissue showing the foot precursor indicated by a red arrow. (d) The average normalized fluorescence intensity along the tissue's morphological axis marked by the foot precursor, as a function of normalized distance from the edge (0 = foot, 1 = head). Averaging is done over ~300 consecutive frames around the morphological transition, and the fluorescence levels are normalized by the maximal value in each frame. The distance from the edge is normalized by the tissue's size and accumulated in 50 equal bins. The dots represent the mean values across 11 tissue samples where the foot precursor marking the morphological axis is clearly identified. The pink-shaded region indicates the standard deviation. See Methods for details. Inset: Examples of BF and fluorescence images showing progressively increasing $Ca^{2+}$ activity along the axis from the foot to the head.

hours preceding the second cut, which are stably transmitted through the large tissue segment to the twin sub-tissue samples derived from it. This inheritance is further supported by the dimensionality reduction analysis presented in the t-SNE map (Fig. 1c, black symbols). The ratio of the average distance between a pair of twin tissue samples (black symbols) to the average distance between non-twin tissues is approximately 0.4, indicating that twin samples typically follow trajectories in morphological space that share similar characteristics. This finding further supports the conclusion that the initial conditions of the tissue, influenced by factors inherited from the parent *Hydra*, significantly affect the regeneration trajectory in morphological space. Moreover, the morphological trajectories of twin tissue fragments, although their size is smaller than the critical size defined in Fig. 2e, resemble those of small near-foot or near-head fragments rather than those of small central fragments, suggesting that factors transmitted from the parent

*Hydra* play a role in shaping the morphological dynamics of these fragments. The transmission of inherited attributes is likely due to the presence of positional cues residing within the tissue. Thus, a tissue's trajectory in morphological space is expected to be influenced by its original position along the parent body axis. Indeed, as shown in Figs. 3c and 3d, small tissue fragments (<450 μm) excised near the head or foot (8 samples each) of the donor *Hydra,* exhibit morphological transitions with characteristics distinctly different from those small tissue fragments taken from the mid-axis (Supplementary Figs. S5 and S6)[8].

**Polarity Establishment and Inherited Calcium-Based Polarity Vector**

A clear indicator of polarity establishment in a regenerating tissue fragment is the early emergence of a foot precursor, which always appears in our experiments before the primary morphological transition into an elongated tube-like shape (Fig. 4c; Supplementary Figs. S2-S6). The tissue's size and its position along the donor *Hydra*'s body are critical factors in determining the timing of polarity establishment via the appearance of a foot precursor. Figs. 4a and 4b present histograms of the approximate upper-limit times for the appearance of the foot precursor for different sample types, showing a clear hierarchy (see also the traces in Supplementary Figs. S2-S3 and S5 S6); short times for mid-axis large tissue fragments ( $600 \pm 300\,\mathrm{min}$ ) and small tissue fragments excised near the original foot ($1000 \pm 550\,\mathrm{min}$ ), and longer times for mid-axis small fragments$\left(1700 \pm 500\,\mathrm{min}\right)$ and for those excised near the original head ( $1700 \pm 300\,\mathrm{min}$ ). These time differences are aligned with the data shown in the t-SNE map in Fig. 1c. The timing of foot appearance is also reflected in the distinct morphological trajectories of the tissue fragments (Figs. 2 & 3). It hints that the processes leading to polarity establishment and morphology are interconnected.

Our previous studies have highlighted the critical role of $Ca^{2+}$ in driving morphological transitions during *Hydra* regeneration, suggesting that $Ca^{2+}$ acts as a dynamic mediator linking these two processes [8, 45]. Utilizing the early emergence of a foot precursor as a clear reference for the body-axis, we measure the $Ca^{2+}$ activity along it using a fluorescent probe[8-10] (for details see Supplementary Note 3). Figure 4d demonstrates that the average $Ca^{2+}$ activity exhibits a clear gradient reflecting the polarity vector; increasing along the axis from the foot precursor towards the future head (see the same trend for individual tissue samples in Supplementary Fig. S7). This trend in the $Ca^{2+}$ activity is clearly observed in the inset images in Fig. 4d.

We next inquire whether such a gradient also exists at earlier stages of the regeneration process, by analyzing the average $Ca^{2+}$ density from the onset within a sampling circle around the tissue's center of mass. The $Ca^{2+}$ density within this circle is estimated at different time points by accumulating the signal in distance bins from 15 distinct pairs of antipodal points on the diameter of the sampling circle and averaging over 50-time frames (50 min). The pair yielding the largest gradient is chosen as a reference polarity axis (see Supplementary Note 3). As shown in Fig. 5b, the fluorescence profiles along diameters away from that associated with the maximal gradient, deviate gradually by first flattening out and eventually flipping direction. Remarkably, our data indicate that a clear $Ca^{2+}$ gradient, in all samples, is already present at the earliest observation point (~2-3 hrs post excision), well before any foot precursor is visible. Comparing this early gradient with the polarity axis measured by the alternative method in Fig. 4d, shows that it is consistently aligned with the axis defined by the foot precursor. As far as we know, this is the first time a continuous gradient reflecting polarity, previously inferred only indirectly through grafting experiments, has been directly observed in *Hydra* [32, 44, 51]. Fig. 5c provides an example where the gradient remains almost constant with time, at least until the tissue begins to elongate significantly, suggesting that the polarity indicated by the $Ca^{2+}$ activity gradient is inherited from the parent *Hydra*. Further support comes from Figure 5d, showing that tissue fragments excised closer to the donor's head display higher initial $Ca^{2+}$ activity than those taken near the foot, consistent with a persistent $Ca^{2+}$ gradient in the adult animal. Over time, however, as the regenerating tissue approaches its main morphological transition, these average $Ca^{2+}$ levels converge toward similar values.

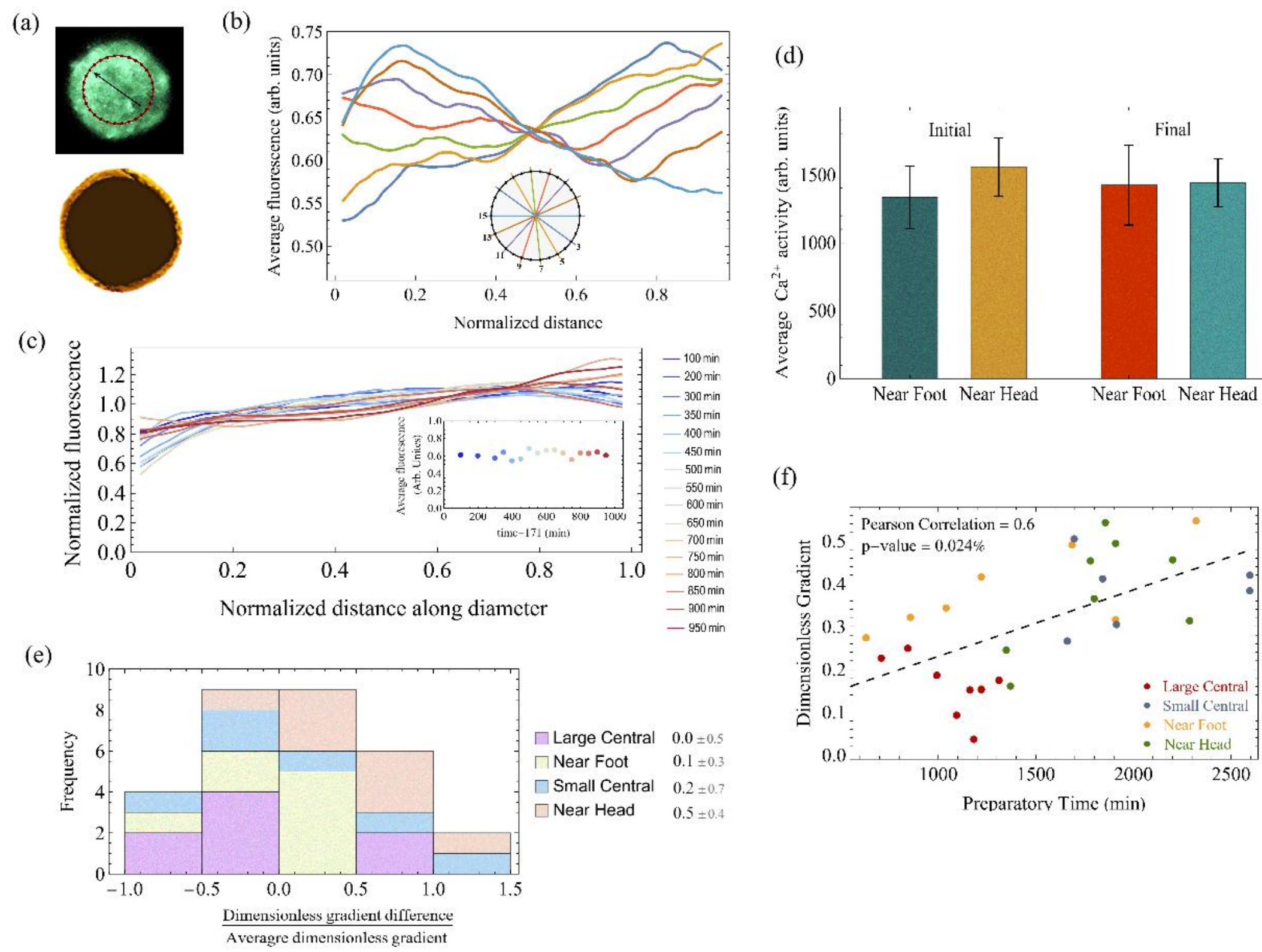

**Figure 5. Early Polarity Inheritance and $Ca^{2+}$ Gradient Evolution.** (a) Representative BF and fluorescence images of a tissue fragment at early stages after folding into a hollow spheroid. The fluorescence signal probes the $Ca^{2+}$ activity. A sampling circle (red), centered at the fragment's center of mass, spanning 70% of its projected diameter, is used to measure the average fluorescence along diameters connecting antipodal vertices (30 total; Supplementary Note 3). The arrow indicates the $Ca^{2+}$ gradient direction deduced from the analysis in (c). (b) Example fluorescence profiles measured along 7 representative diameters. Each profile results from averaging over 50 frames (50 min). The antipodal vertices with the largest absolute gradient define the axis. (c) Normalized $Ca^{2+}$ distributions (mean = 1) at 16 time points during the "preparatory" interval-the time period prior to the morphological transition. The analysis is done as in (b), and the distribution presented at each time point, marked at the color legend (representing an average over 50 frames at the time point), is the one showing the maximal gradient (the marked time is from the onset of measurement, 2-3 hrs post excision). The measured $Ca^{2+}$ density exhibits a persistent gradient that emerges very early following the folding of the tissue fragment, indicating inherited polarity that is imprinted from the onset in the $Ca^{2+}$ signal. The inset shows the average fluorescence versus time over the same interval. (d) A comparison of mean $Ca^{2+}$ activity in near-foot and near-head samples at the beginning ("initial") and end ("final") of the preparatory period. (e) A stacked histogram of the change in dimensionless gradient (from initial to final), normalized by its average value during the "preparatory time". Different colors correspond to distinct initial conditions. The dimensionless $Ca^{2+}$ gradient is defined as $L\overline{(\partial\rho/\partial x)}/\bar{\rho}$ , where $\rho(x)$ is the $Ca^{2+}$ density along the identified polar axis, overbar denotes a spatial average, and L is the characteristic tissue size. (f) A scatter plot of the final dimensionless gradient against the "preparatory time", revealing a positive correlation (Pearson coefficient = 0.6; p-value = 0.024%).

While Fig. 5c shows a remarkable stability over time, both in the $Ca^{2+}$ gradient (main) and in the average $Ca^{2+}$ activity (inset), this is not always the case. Generally, the initial gradient can differ from the one measured near the main morphological transition. Fig. 5e presents a stacked histogram quantifying these gradient changes, grouped by various tissue types. Near-head samples typically exhibit an increase in the gradient during the "preparatory time" - the time interval before the primary morphological transition, while near-foot and large central tissues remain comparatively stable. Small central fragments also show an increase, though to a lesser degree than near-head pieces. Finally, Fig. 5f demonstrates that the dimensionless gradient near the morphological transition time correlates positively with the duration of the "preparatory time".

**Mechanochemical Integration of Polarity and Morphogenesis**

Morphogenesis during regeneration can be viewed by analogy as the motion of an overdamped particle subjected to noise, navigating through a slowly evolving, high-dimensional morphological landscape[8, 45]. Each regenerating tissue fragment inherits positional information from its original axial position. These inherited factors specify initial conditions which in turn can shape the energy landscape itself and determine the subsequent distinctive trajectories, leading to a canalized morphogenesis[52]. Small mid-axis fragments typically exhibit rapid morphological transitions, analogous to a particle undergoing noise-driven activation over an energy barrier separating two distinct morphological states: a nearly spherical configuration and an elongated, tube-like shape[8, 45]. In contrast, larger fragments or those taken from regions near the head or foot carry positional cues that reshape the morphological energy landscape, rendering it more rugged and structured by multiple shallow minima. As a result, morphological changes proceed more gradually or in discrete steps, with the tissue occasionally residing in transient, "metastable" configurations. The twin-fragment experiments provide clear evidence of canalization, as tissue fragments with similar initial conditions consistently follow similar regenerative trajectories.

Other processes in regenerating tissue fragments also reflect the underlying positional cues. Polarity patterns reflected in the $Ca^{2+}$ activity clearly indicate sensitivity to such cues. Near-foot and large mid-axis fragments typically establish stable $Ca^{2+}$ gradients early on, whereas near-head fragments begin with weaker gradients that gradually intensify. Also, the timing of foot precursor emergence is aligned with positional cues: appearing earlier in near-foot and large mid-axis fragments, and later in near-head or small mid-axis ones. Yet, similar timings of foot precursor formation do not always lead to identical morphological outcomes, highlighting the complexity of the morphological landscape.

Despite years of research, the processes responsible for the inherited polarity gradient and the positional information affecting regeneration remain elusive. Our data suggest that this inherited information cannot solely come from biosignals emerging from the head organizer. Regardless of the precise underlying processes, their effect on the coupling between polarity establishment and morphological transition can be rationalized by introducing a simple mechanochemical toy model, which provides a basis for a more elaborated future framework. In its basic form, the model consists of three key components that govern the likelihood of the tissue adopting different morphological trajectories: (i) a tissue bending energy that favors shapes close to a preferred curvature, (ii) a field representing $Ca^{2+}$ activity characterized by spatial continuity and local excitability, and (iii) a local coupling between $Ca^{2+}$ activity and the tissue's curvature.

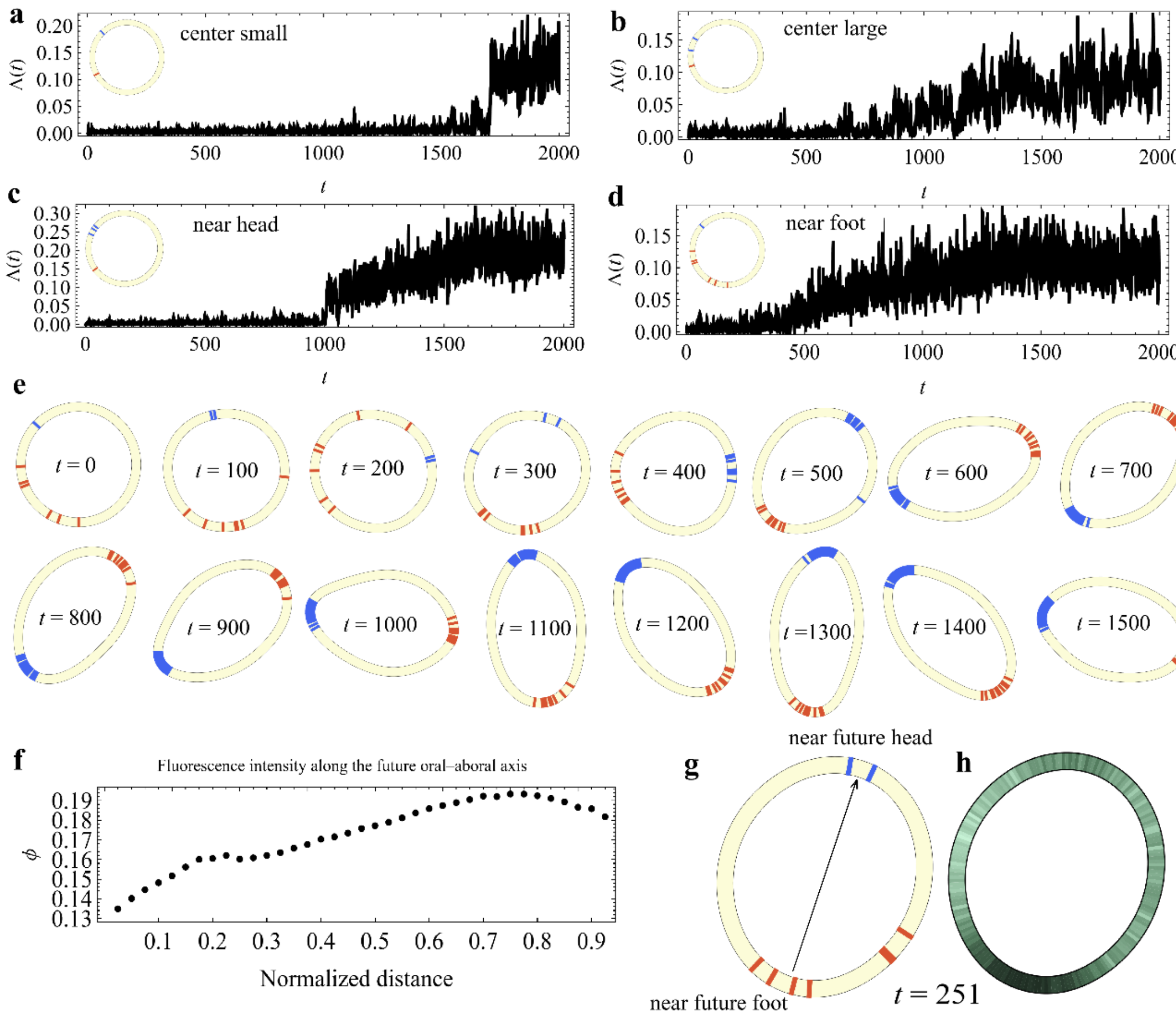

**Figure 6. Simulated regeneration dynamics in the toy model.** (a-d) Representative simulations showing the temporal evolution of the shape parameter in a 1-D toy model of a ring tissue, for different initial tissue types representing: (a) small fragments from the central gastric region, (b) large tissues from the central gastric region, (c) small tissues excised near the head, and (d) small tissues excised near the foot. Small and large fragments correspond to rings with a small or large number of "cells," respectively. The initial state of each fragment is specified by the spatial distribution of F-type (red) and H-type (blue) cues along the ring, as well as by the initial coupling strength between the tissue line curvature and the $Ca^{2+}$ field (for details, see Supplementary Note 4).(e) Example of the dynamics in a near-foot tissue, showing the evolution of tissue shape and positional cues at several time points. The snapshots illustrate the morphological evolution and the segregation of positional cues into opposite poles. (f) Profile of the scalar field $\phi$ (proxy for $Ca^{2+}$ activity), plotted as a function of distance from the center of mass of F-type cells and averaged over the first 200 simulation frames. (g) A representative intermediate configuration of the system. The arrow indicates the polarity axis. (h) An instantaneous density map of $\phi$ illustrating its spatial gradient. For clarity, color brightness is shown on an exponential intensity scale.

Together, these elements define an effective morphological landscape and recapitulate the experimentally observed morphological transitions in small mid-axis tissue fragments[8, 45, 53].

However, in this baseline model the elongation axis emerges spontaneously through fluctuations. In contrast, experimental observations indicate that polarity cues are present from the onset of regeneration and play a decisive role in determining the morphological axis. This discrepancy motivates an extension of the model to explicitly include inherited positional information, which in turn determines the axis alignment according to the original donor's polarity.

In line with our experimental findings, we introduce two positional cues: foot-associated (F-type) and head-associated (H-type). These cues evolve under stochastic cellular-automaton rules and interact through biosignals for which they serve as sources. During the preparatory phase and the subsequent morphological transition, they spontaneously cluster at opposite ends of the tissue, thereby establishing polarity (Fig. 6e). Locally, they modulate bending stiffness and $Ca^{2+}$ activity: biosignals from H-type cues enhance $Ca^{2+}$ activity and reduce the bending modulus, whereas F-type cues suppress $Ca^{2+}$ activity. Together, these effects generate a stable $Ca^{2+}$ gradient that breaks symmetry and directly influences tissue shape. The detailed implementation of the above model is described in Supplementary Note 4.

Figures 6a–d show representative time courses of the model shape parameter for four initial conditions: small fragments from the mid-body, large mid-body tissues, and small tissues excised near the head or near the foot (additional examples in Supplementary Figs. S9-S12). The insets display the initial distributions of positional cues; for example, near-foot tissues begin with several F-type (red) cells and only a single H-type (blue) cell. The simulations reproduce the experimental trends: an abrupt morphological transition for small mid-body tissues, and a gradual, more complex transition for the other cases (compare Fig. 6a,b with Fig. 2a,b, and Fig. 6c,d with Fig. 3c,d). Figure 6e illustrates segregation of the positional cues to opposite poles, highlighting the interplay between morphological change and polarity establishment. Figure 6f shows the average $\phi$ field (a proxy for $Ca^{2+}$ activity), which exhibits a gradient similar to that observed experimentally (Fig. 4d).

**Conclusions**

Our findings portray *Hydra* regeneration as a multi-canalized process, in which polarity and morphological-axis formation progress in parallel and rely on each other - reflecting the dual nature of the developmental body axis. Local positional information shapes both a complex morphological landscape and polarity cues. These, in turn, are mirrored and likely influenced by the spatial distribution of the $Ca^{2+}$ activity, which can feed back into the tissue's mechanical and morphological dynamics. Moreover, our data suggest the presence of two primary signal sources in the mature *Hydra*, one near the head and another near the foot, rather than a single axial gradient originating from a single source in the head as often assumed. Taken together, these insights call for a revision of the conventional view regarding the head organizer and its biochemical signals as the major actors determining polarity and morphogenesis. Instead, multiple inherited factors -biochemical, mechanical, and electrical - concurrently shape the morphological landscape and the polarity signals that guide regeneration. Our simplified theoretical model illustrates these conclusions.

Our data highlight $Ca^{2+}$ as a central regulator of *Hydra* regeneration and a dynamic mediator linking the polarity and morphological facets of body-axis formation. While $Ca^{2+}$ clearly shapes tissue mechanics via actomyosin force generation, its coupling to biosignaling remains elusive. Most of the attention in *Hydra* regeneration is focused on the canonical *Wnt*/*b-catenin* process and the head organizer as a major morphogen source. However, our finding that $Ca^{2+}$ plays a central role in *Hydra* regeneration motivates probing additional pathways, such as noncanonical *Wnt* signaling, which can couple directly to $Ca^{2+}$. Consistent with this view, bud and tentacle formation in *Hydra* have been linked to noncanonical *Wnt* signaling processes[54]

At this stage, hierarchical causal relations among the various regeneration processes cannot be established. However, regardless of whether $Ca^{2+}$ is a primary causal factor, our data show that its profile, dynamics, and spatial distribution are shaped by positional information from both head and foot. Given the electrically excitable nature of the epithelium, these dynamics arise from processes distributed across the tissue rather than from localized sources. In this sense, $Ca^{2+}$ is not a classical morphogen. This points to a new framework for pattern formation, distinct from the traditional Turing-type reaction–diffusion models.

## Materials and Methods

*Experimental Methods*

Experiments are carried out with a transgenic strain of *Hydra Vulgaris* (*AEP*) carrying a GCaMP6s fluorescence probe for $Ca^{2+}$ (see Refs. [8-10]for details of the strain). Animals are cultivated in a *Hydra* culture medium (HM; 1mM NaHCO3, 1mM CaCl2, 0.1mM MgCl2, 0.1mM KCl, 1mM Tris-HCl pH 7.7) at 18°C. The animals are fed every other day with live *Artemia nauplii* and washed after ~4 hours. Experiments are initiated ~24 hours after feeding. Small tissue fragments are excised from different regions along the axis of a mature *Hydra*, close to the head, close to the foot and at the center. A thin ring is first cut from each of these regions and is further cut into (usually) 4 small fragments. These tissue fragments are incubated in a dish for ~3 hrs to allow their folding into spheroids prior to transferring them to the experimental sample holder. Large tissue fragments are prepared by first cutting a wide ring from the central region of a mature *Hydra*, and then further cutting it into two large fragments. These tissue fragments are incubated in a dish for ~4-5 hrs to allow their folding into spheroids prior to transferring them into the experimental sample holder. For the twin experiments, large tissue fragments are allowed to fold in the *Hydra* medium for ~5 hrs and then cut again into two segments which are allowed to re-fold for ~3 hrs before being transferred to the experimental sample holder. The twin tissue samples are recorded under the same conditions.

The experimental setup is similar to the one described in Ref[8-10] without the external electrodes. In all the experiments, spheroid tissues are placed within wells of ~1.3 mm diameter made in a strip of 2% agarose gel (Sigma) to keep the regenerating *Hydra* in place during time-lapse imaging. The tissue spheroids, even the large ones, are free to move within the wells. The agarose strip, containing 15 wells, is fixed on a transparent plexiglass bar of 1 mm height, anchored to a homemade sample holder. A channel on each side of the sample wells allows a continuous medium flow. A peristaltic pump (IPC, Ismatec, Futtererstr, Germany) flows the medium continuously from an external reservoir (replaced at least once every 24 hrs) at a rate of 170 ml/hr into two channels around the samples. The medium covers the entire preparation, and the volume in the bath is kept fixed throughout the experiments by pumping the medium out from 4 holes determining the fluid's height. The continuous medium flow ensures stable environmental conditions. All the experiments are done at room temperature.

*Microscopy*

Time-lapse bright-field and fluorescence images are taken by a Zeiss Axio-observer microscope (Zeiss, Oberkochen Germany) with a 5× air objective (NA=0.25) and acquired by a CCD camera (Zyla 5.5 sCMOS, Andor, Belfast, Northern Ireland). The sample holder is placed on a movable stage (Marzhauser, Germany), and the entire microscopy system is operated by Micromanager, recording bright-field and fluorescence images at 1-minute intervals. The 1 min resolution is chosen on the one hand to allow long experiments while preventing tissue damage throughout the experiments and, on the other hand, to enable recordings from multiple tissue samples.

*Data Analysis*

For the analysis, images are reduced to 696x520 pixels (~2.5 µm per pixel) using ImageJ. Masks depicting the projected tissue shape are determined for a time-lapse movie using the bright-field (BF) images by a segmentation algorithm described in [55] and a custom code written in Matlab. Shape analysis of regenerating *Hydra*'s tissue is done by representing the projected shape of the tissue by polygonal outlines using the Celltool package developed by Zach Pincus [56]. The polygons derived from the masks provide a series of (x,y) points corresponding to the tissue's boundary. Each series is resampled to 30 points, which are evenly spaced along the boundary. The fluorescence analysis is done on images reduced to the same size as the bright-field ones (696x520 pixels).

*Simulations***.**

We model the tissue as a one-dimensional closed ring of $N$ sites (small and large fragments: $N$=200 and $N$=400, respectively), each associated with a position $\boldsymbol{R}_i(t)$ along a closed tissue curve whose local curvature is $H_i(t)$. On this ring we track (i) a scalar activation field $\phi_i(t)$ representing local $Ca^{2+}$ activity, (ii) two diffusive biochemical signals $h_i(t)$ and $f_i(t)$ (*H*- and *F*-signals), and (iii) binary *H*- and *F*-type "cells" $n_{H,i}$, $n_{F,i} \in \{0,1\}$, The signals $h$ and $f$ obey coupled reaction–diffusion equations with production by *H*-/*F*-cells, linear decay, and mutual annihilation. The $Ca^{2+}$ field represented by $\phi$ evolves as a noisy gradient flow in a signal- and curvature-modulated tilted double-well potential $U(\phi_i;h_i,f_i,H_i)$ and constrained to be non-negative. *H*- and *F*-cells undergo biased random walks along the ring (up/down the $h$-gradient) and stochastic birth–death processes controlled by soft thresholds in $h$ and $f$. The tissue boundary $\boldsymbol{R}_i(t)$ is updated by Monte Carlo moves that deform the curve at fixed area, with an energy that penalizes deviations from a preferred curvature and couples curvature to $\phi$. Diffusion terms are integrated using a spectral scheme; other fields are advanced with an explicit Euler–Maruyama update; geometric moves are accepted with a Metropolis criterion. A detailed description of the model is provided in Supplementary Note 3, and all parameter values, initialization protocols, and the full Mathematica code are given in the Supplementary Software.

**Acknowledgements:**

We thank Omri Gat, Kinneret Keren, Shani Maoz, and Omri Wurtzel for useful discussions and comments on the manuscript. EB thanks Liora Garion for technical help. This work was supported by a grant (EB) from the Israel Science Foundation (Grant No. 1638/21)).

**List of Supplementary Materials:**

Supplementary Figs. S1 to S13
Supplementary Notes 1-5
References
Supplementary software

Supplementary Material for

# The Dual Nature of Body-Axis Formation in *Hydra* Regeneration: Polarity-Morphology Concurrency

Oded Agam[1] and Erez Braun[2]

[1]The Racah Institute of Physics, Edmond J. Safra Campus, The Hebrew University of Jerusalem, Jerusalem 9190401, Israel.

[2] Department of Physics and Network Biology Research Laboratories, Technion-Israel Institute of Technology, Haifa 32000, Israel.

**This PDF file includes:**

Supplementary Figs. S1 to S13
Supplementary Notes 1-5
References
Supplementary Software

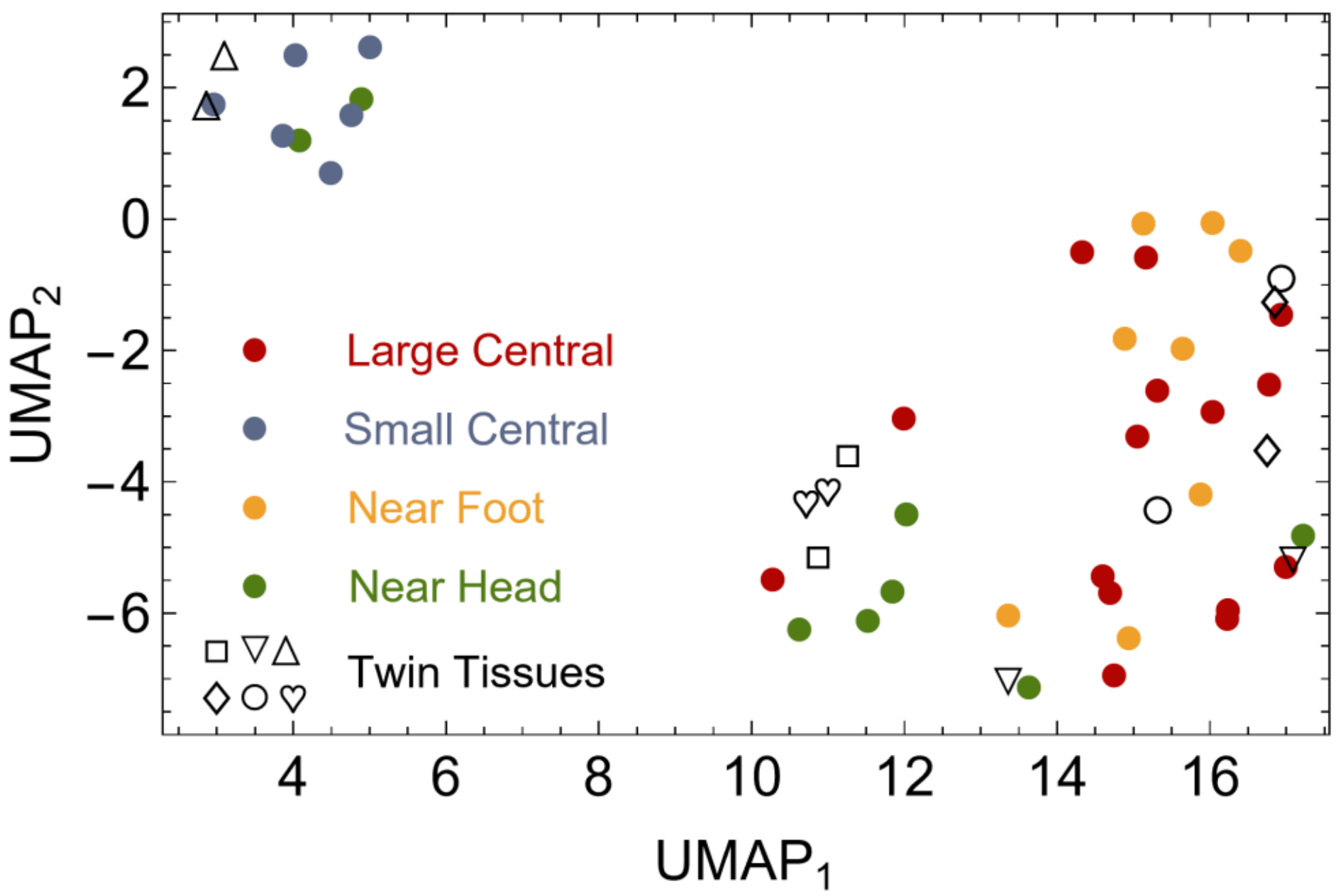

**Fig. S1. Dimensionality reduction by UMAP.** The same experimental protocols and corresponding morphological evolution classification used in Fig.1, utilizing the UMAP method instead of t-SNE.

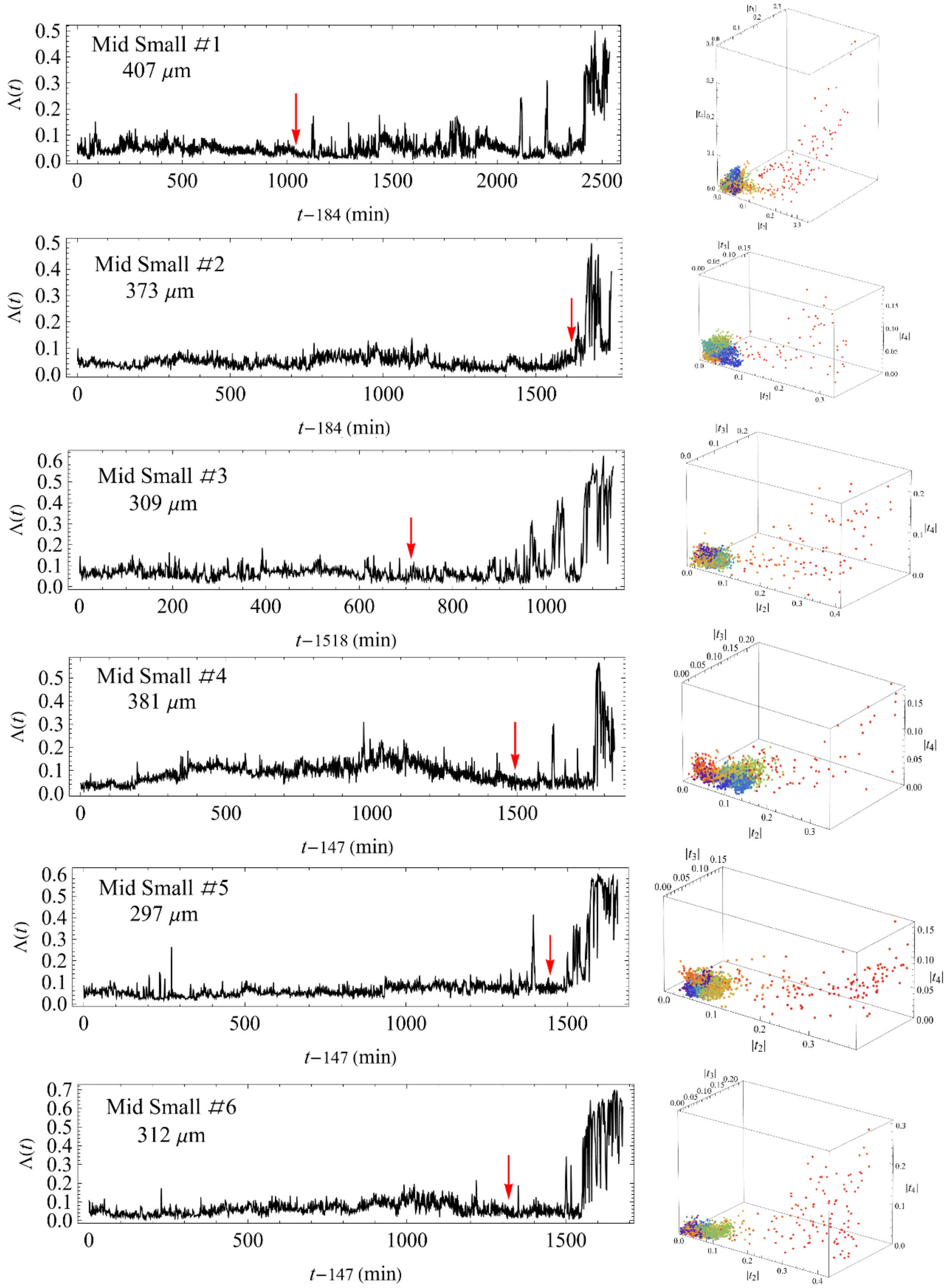

**Fig. S2 Time traces and harmonic moments of individual small tissue samples.** (Left) The shape parameter traces of individual tissue samples. (Right) The regeneration trajectories in the reduced morphological space, defined by the absolute values of the leading harmonic moments, with a rainbow color code representing time (purple for early times, red for later stages). The data shown is for small tissue samples excised from the mid-axis region of the parent *Hydra*. The red arrows mark the time point of the emergence of a foot precursor that can be identified in the bright-field microscopy images. Time is measured from the first imaging time point, 2-3 hrs after excision of the tissue fragment.
`

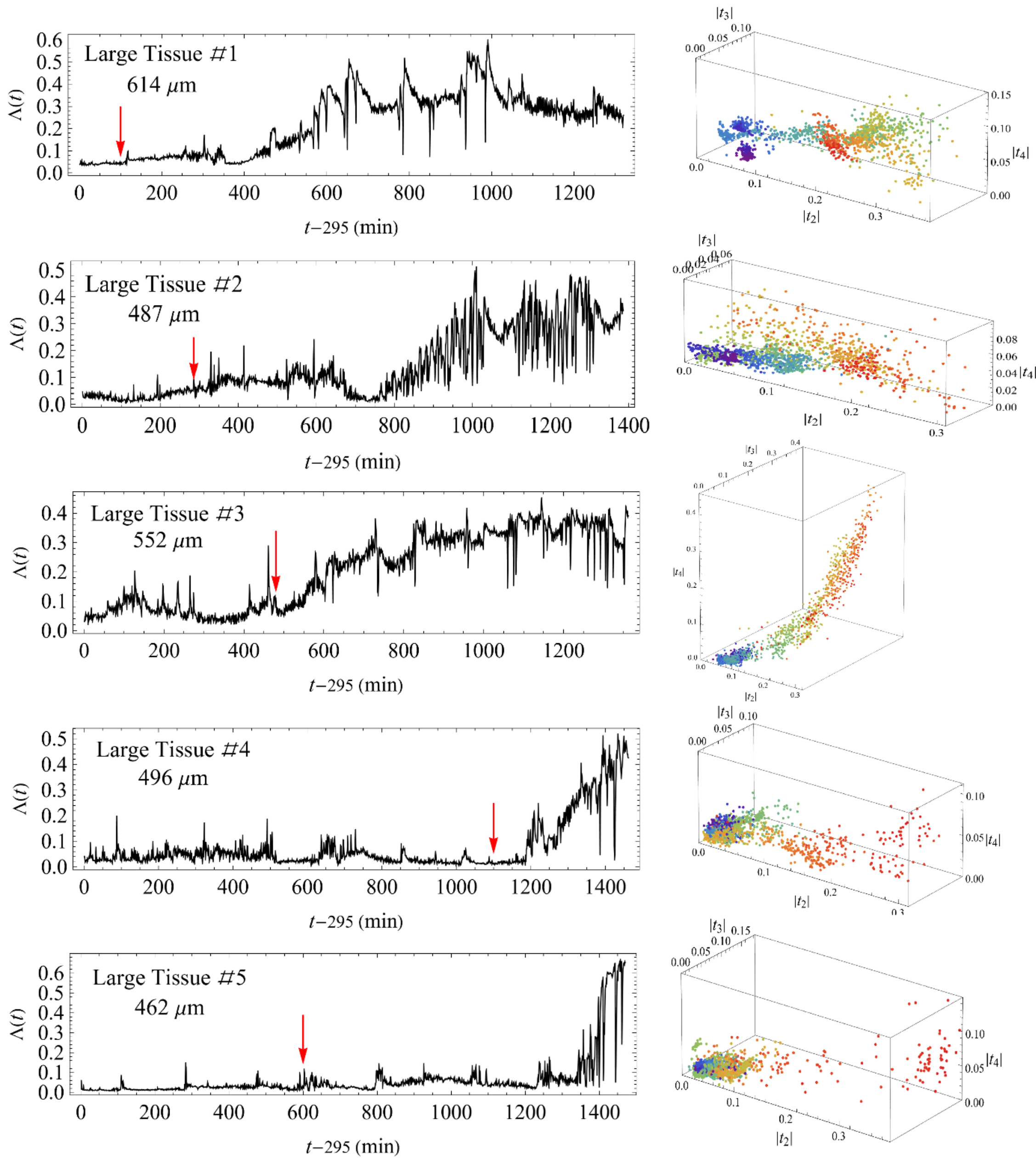

**Fig. S3 Time traces and harmonic moments of individual large tissue samples.** The same as in Fig. S2 for large tissue samples excised from the mid-axis region of the parent *Hydra*.

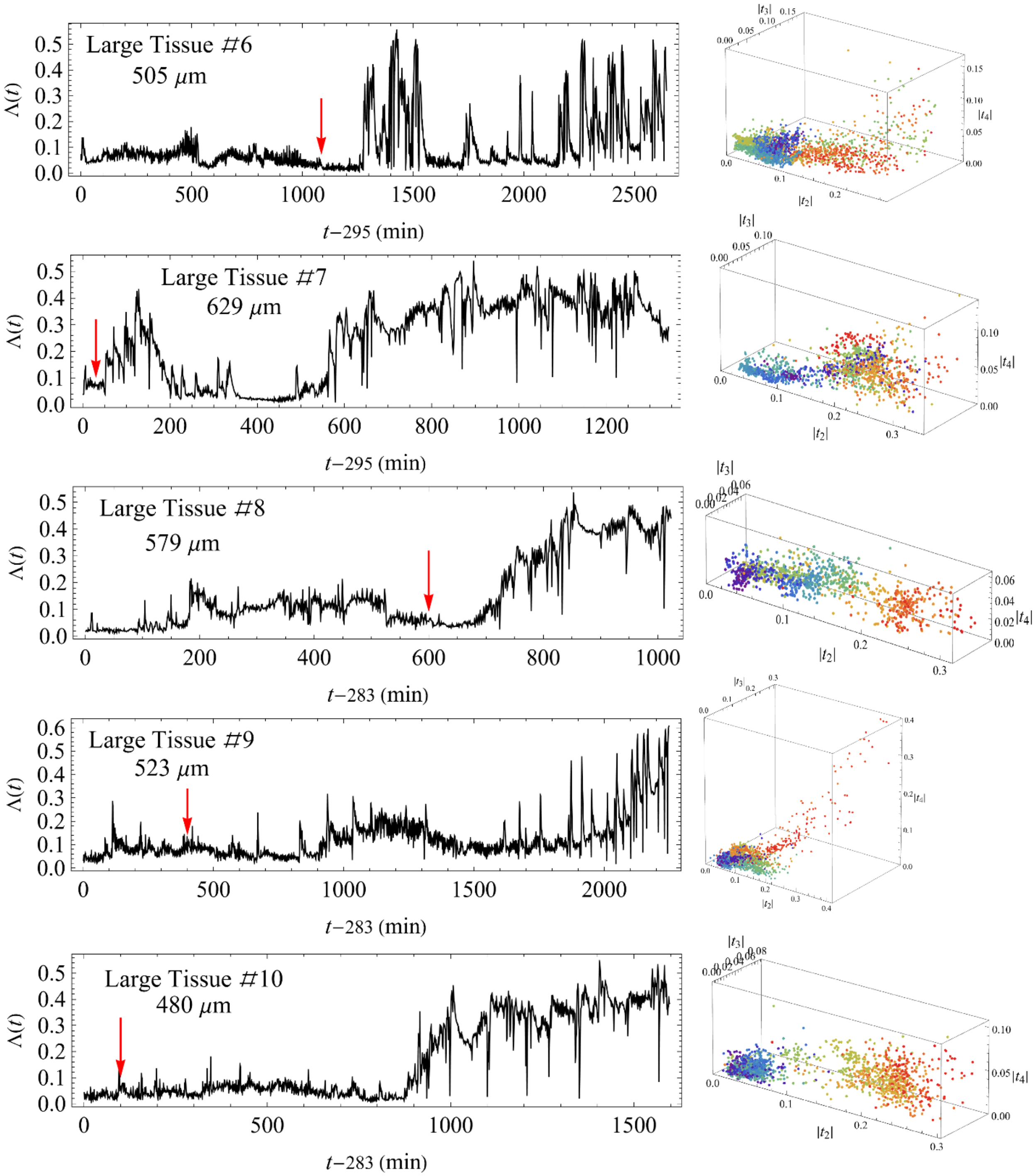
Large Tissue #6
505 μm
Λ(t)
t−295 (min)
Large Tissue #7
629 μm
Λ(t)
t−295 (min)
Large Tissue #8
579 μm
Λ(t)
t−283 (min)
Large Tissue #9
523 μm
Λ(t)
t−283 (min)
Large Tissue #10
480 μm
Λ(t)
t−283 (min)
|t2|
|t3|
|t4|

**Fig. S3 (Cont.)**

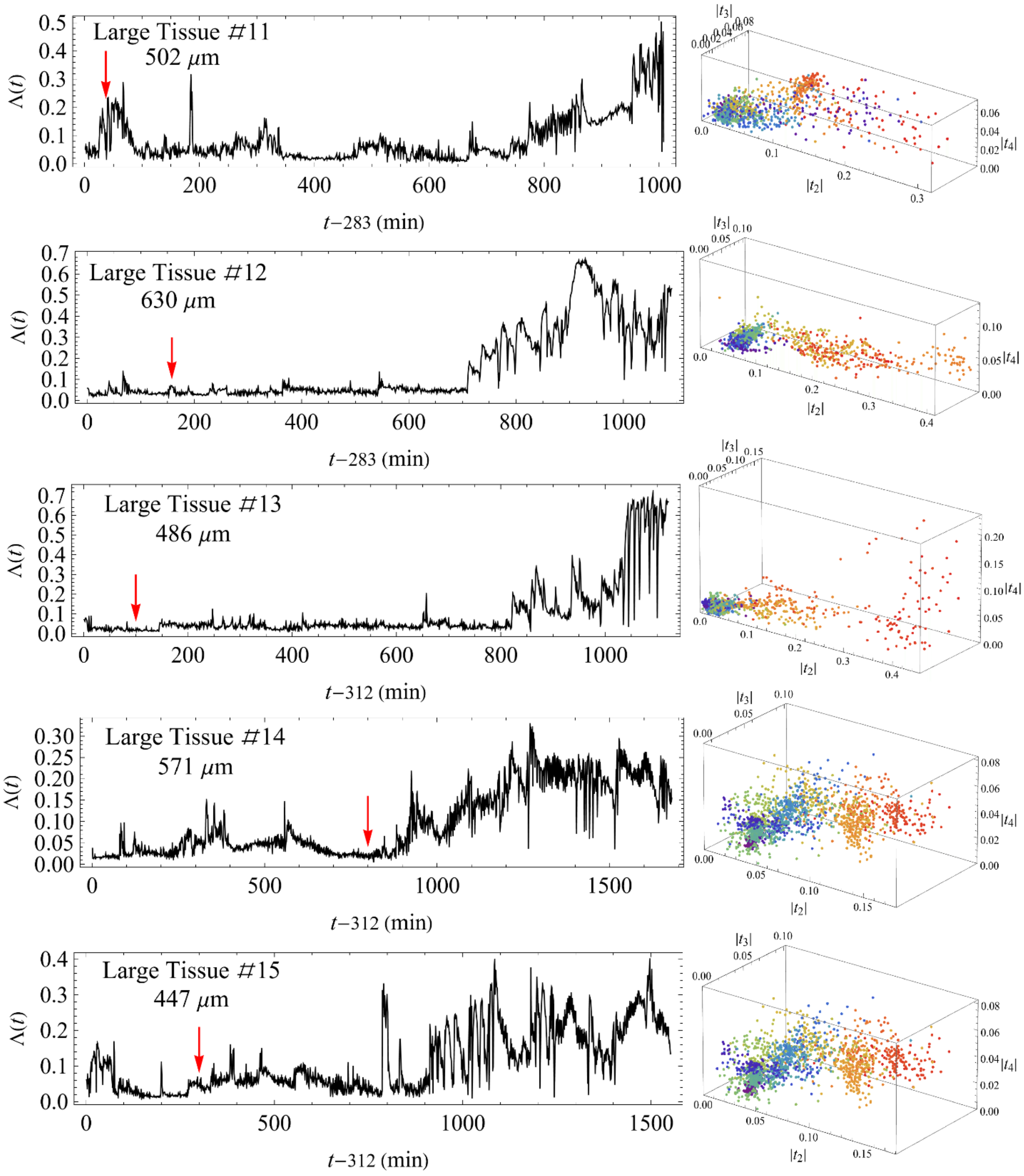

Large Tissue #11
502 μm
Λ(t)
t−283 (min)
Large Tissue #12
630 μm
Λ(t)
t−283 (min)
Large Tissue #13
486 μm
Λ(t)
t−312 (min)
Large Tissue #14
571 μm
Λ(t)
t−312 (min)
Large Tissue #15
447 μm
Λ(t)
t−312 (min)
|t2|
|t3|
|t4|

**Fig. S3 (Cont.)**

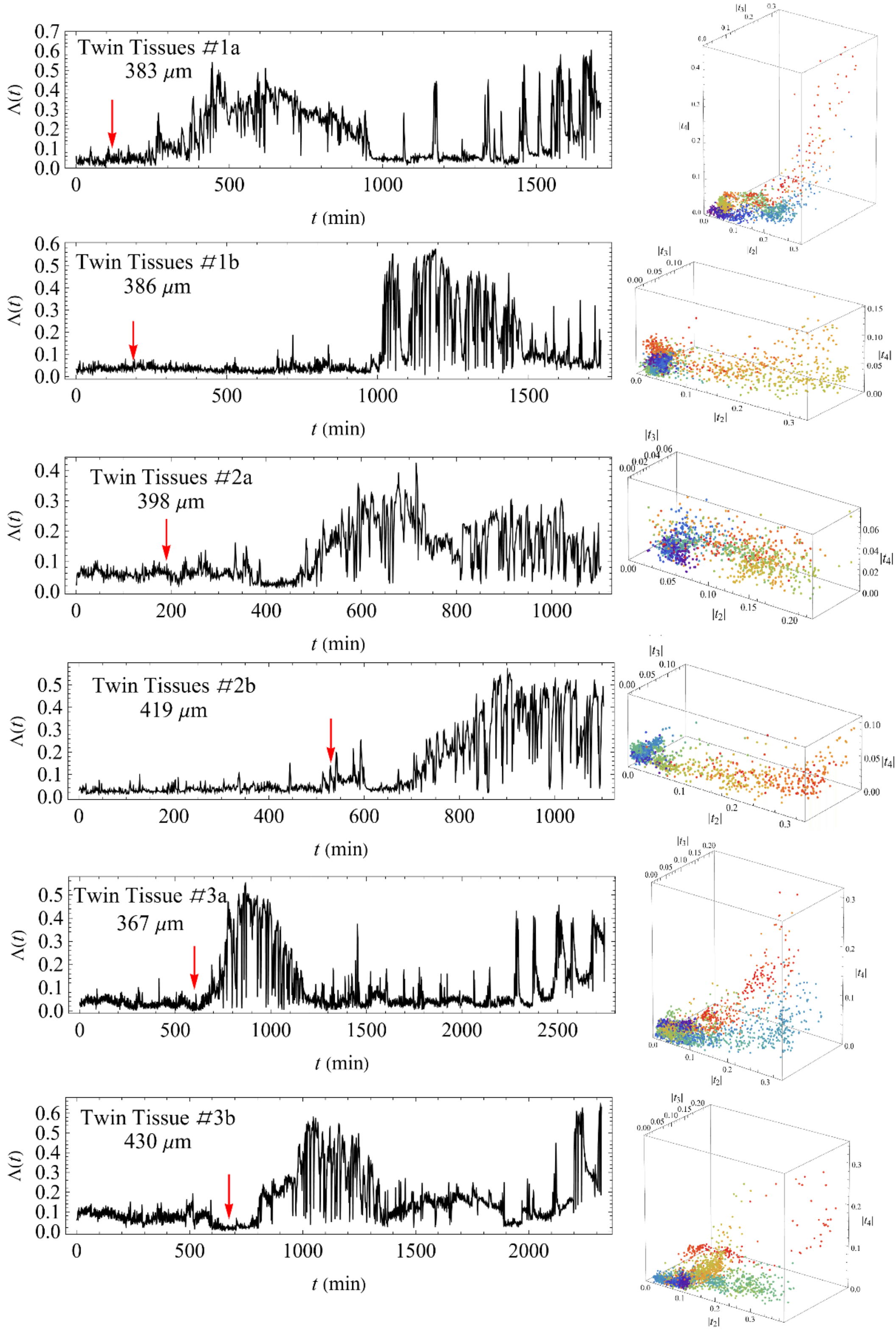

**Fig. S4. Time traces and harmonic moments of twin tissue samples.** The same as in Fig. S2 for twin samples. Each pair of samples is derived from a single large tissue fragment excised from the mid-axis of the parent *Hydra* (see Methods).

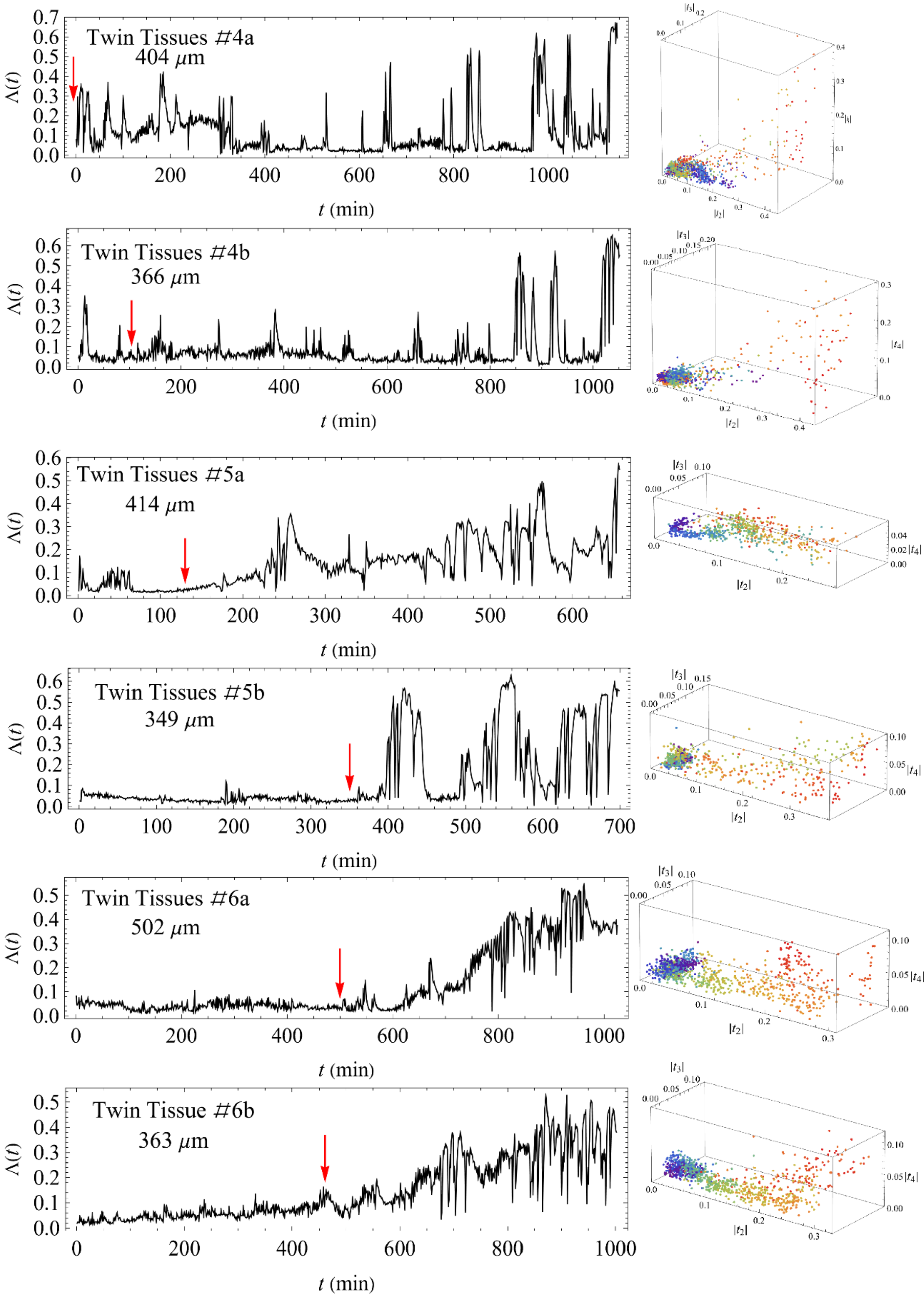

Twin Tissues #4a
404 μm
Λ(t)
t (min)
Twin Tissues #4b
366 μm
Twin Tissues #5a
414 μm
Twin Tissues #5b
349 μm
Twin Tissues #6a
502 μm
Twin Tissue #6b
363 μm

**Fig. S4. (Cont.)**

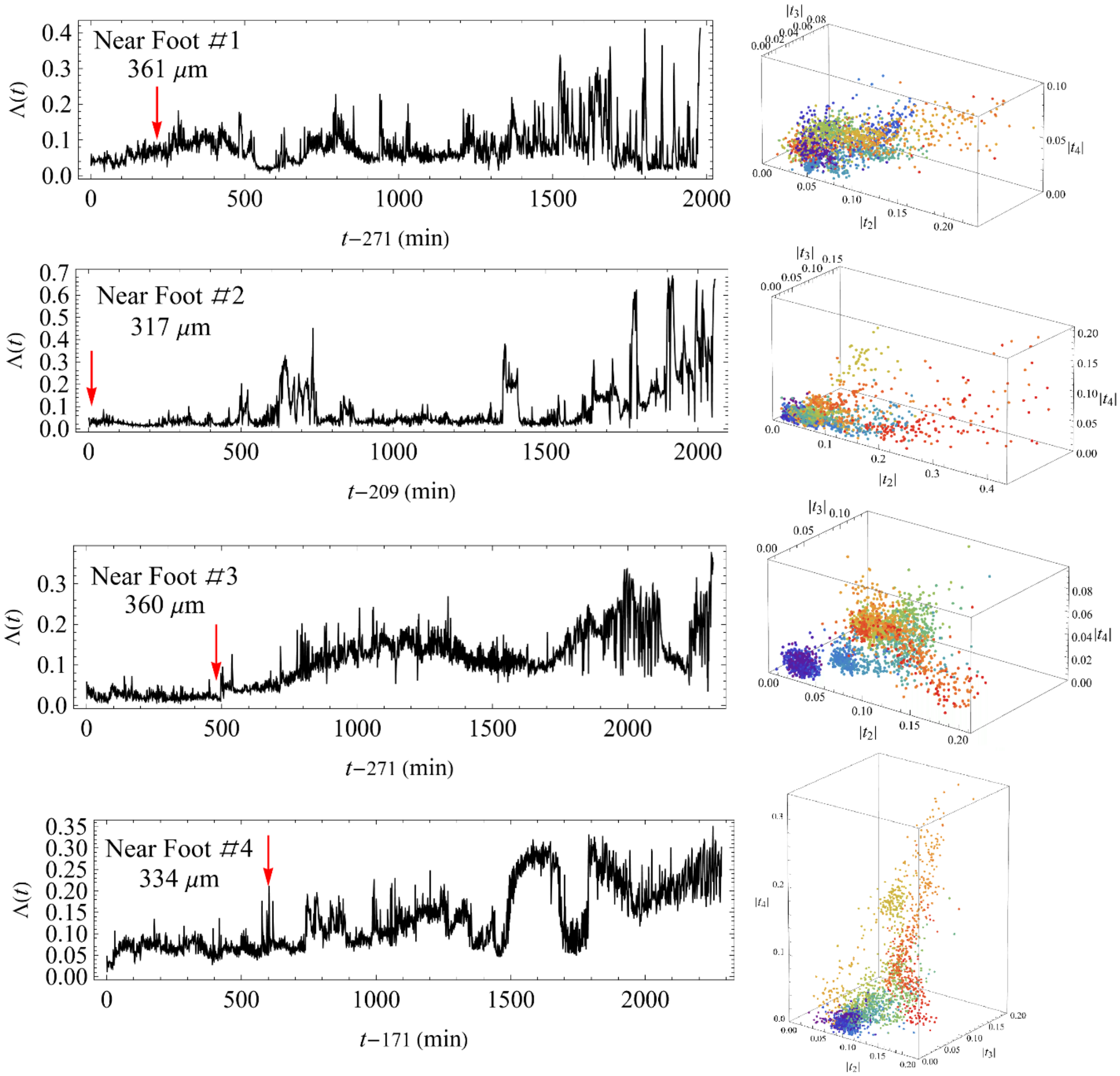

**Fig. S5. Time traces and harmonic moments of individual near-foot tissue samples.** The same as in Fig. S2 for small tissue samples excised from regions near the foot of the parent *Hydra*.

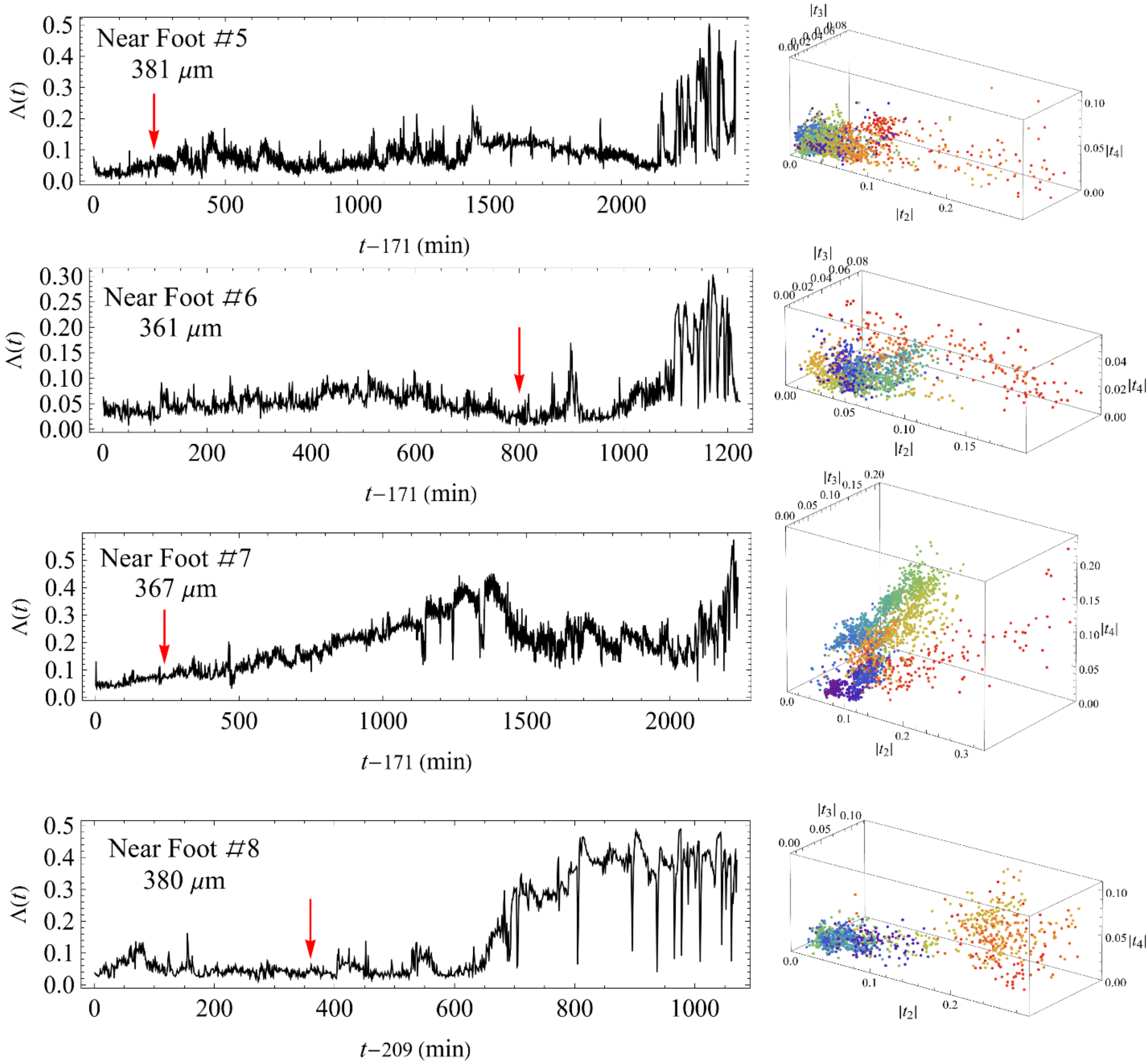

Near Foot #5
381 μm
Λ(t)
t−171 (min)
|t3|
|t2|
|t4|
Near Foot #6
361 μm
Λ(t)
t−171 (min)
Near Foot #7
367 μm
Λ(t)
t−171 (min)
Near Foot #8
380 μm
Λ(t)
t−209 (min)

**Fig. S5. (Cont.)**

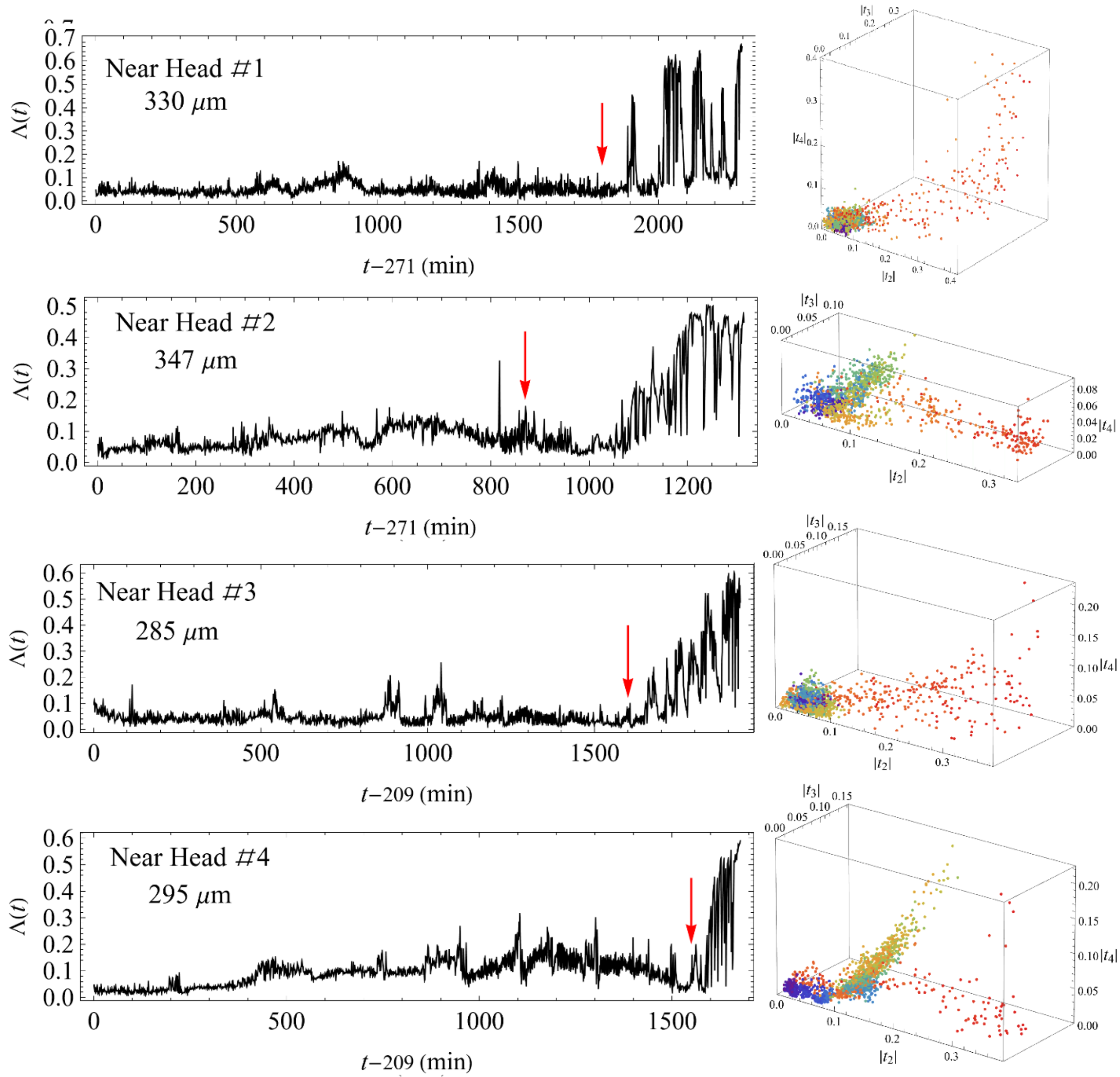

**Fig. S6. Time traces and harmonic moments of individual near-head tissue samples.** The same as in Fig. S2 for small tissue samples excised from regions near the head of the parent *Hydra*.

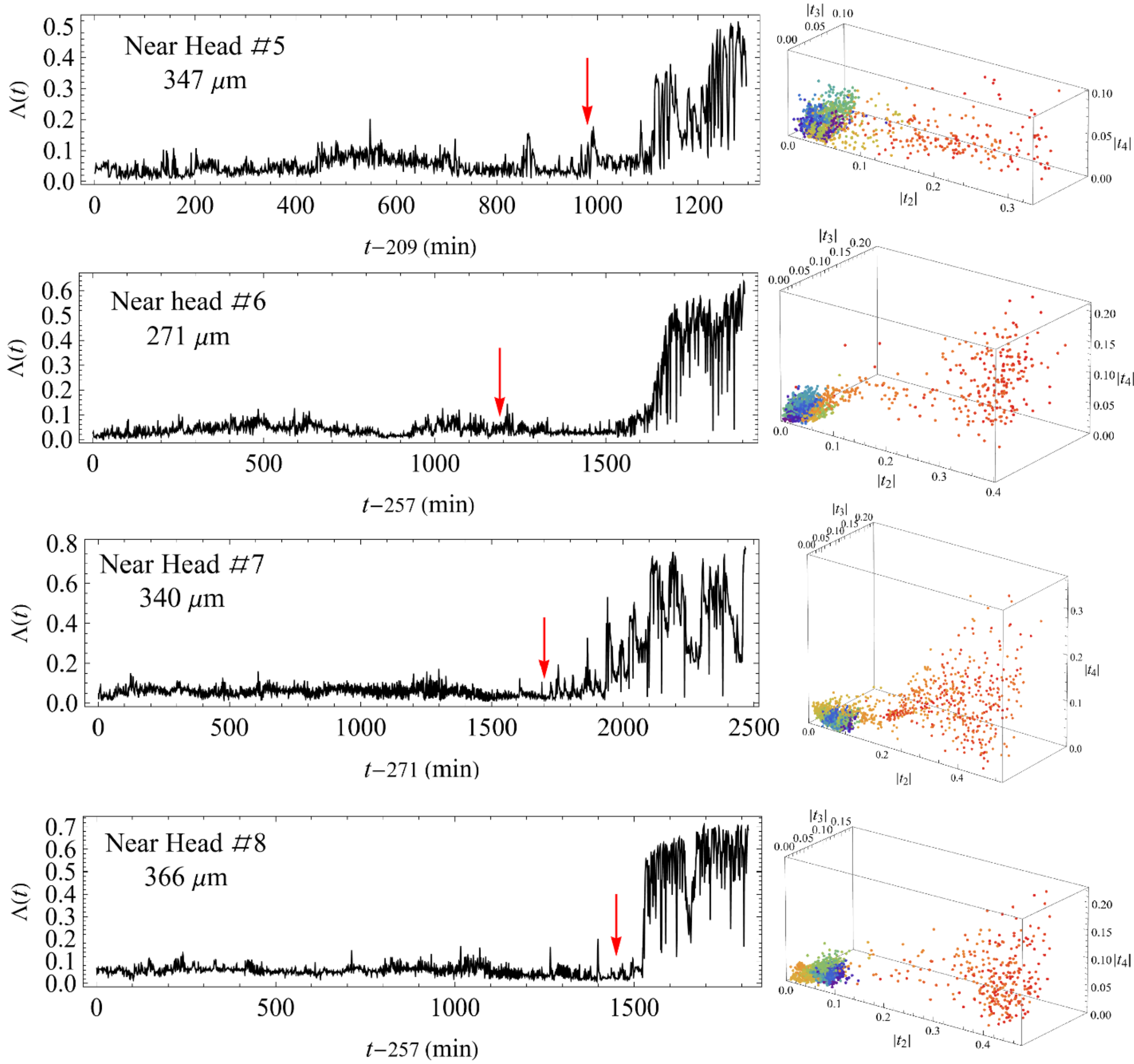
Near Head #5
347 μm
Λ(t)
t−209 (min)
|t3|
|t2|
|t4|
Near head #6
271 μm
Λ(t)
t−257 (min)
Near Head #7
340 μm
Λ(t)
t−271 (min)
Near Head #8
366 μm
Λ(t)
t−257 (min)

**Fig. S6 (Cont.).**

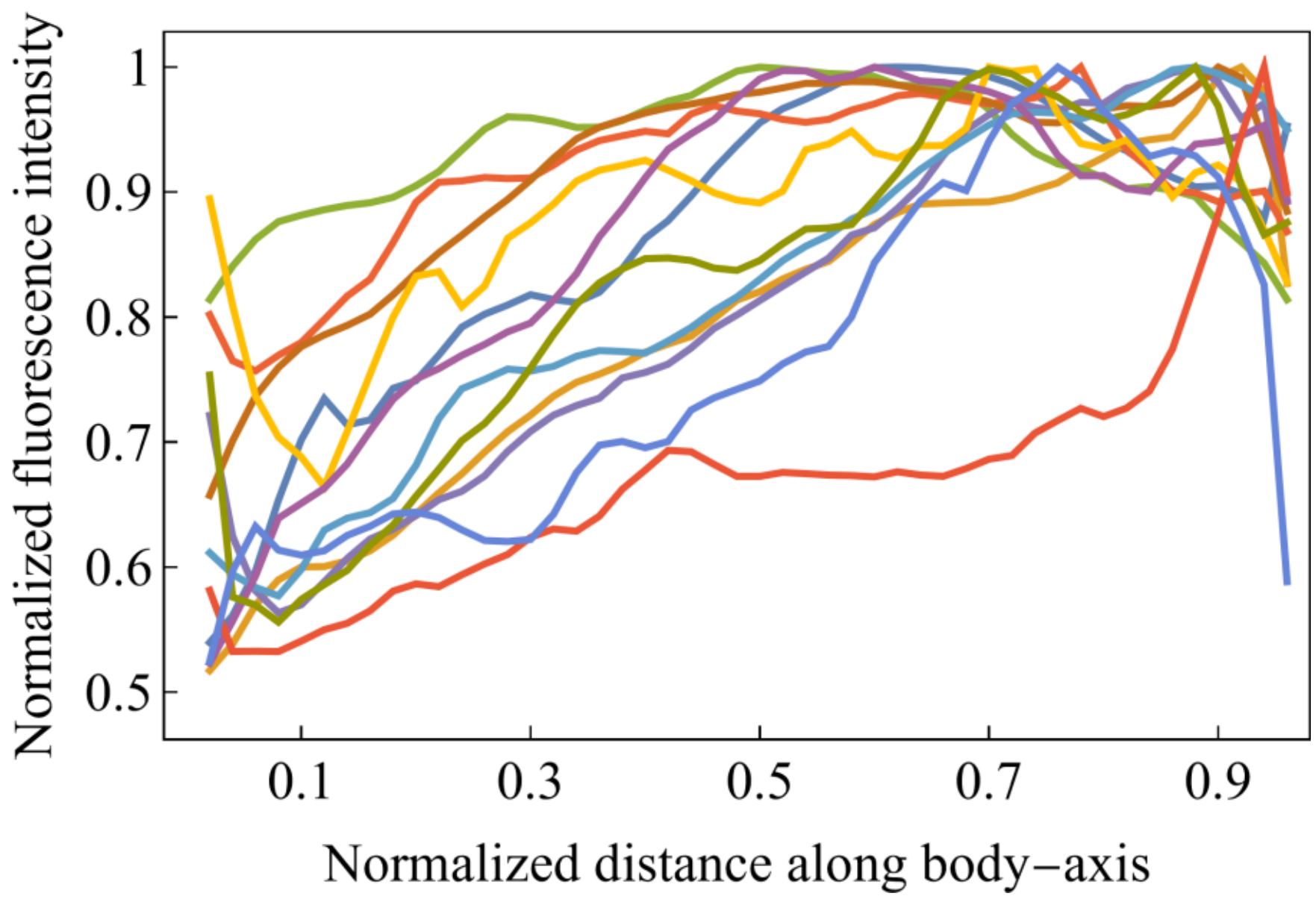

**Fig. S7. Normalized $Ca^{2+}$ average activity along the body axis.** The figure displays curves of the average fluorescence signal—proportional to $Ca^{2+}$ activity—normalized by its maximum value and plotted against the normalized distance (fraction of the full axis length) from the foot precursor. Data from 12 different tissue samples, covering all types of initial conditions, are shown. Note the clear gradient measured for all samples. See Supplementary Note 3 for calculation details.

## Supplementary Note 1: Harmonic Moment Analysis for Tissue Morphology

In this note, we describe how the two-dimensional projection of the tissue at any given time can be viewed as a vector in an infinite-dimensional space [1]. The tissue-projected image can be approximated as a closed, non-intersecting curve in a two-dimensional plane. One way to span the morphological space is by using the area enclosed by the curve, along with the external harmonic moments defined in Eq. (1). The area integral for these moments can be converted into a contour integral using Green's theorem:

$$t_n = -\frac{1}{2\pi n i}\oint dz \frac{z^*}{z^n}, \quad n = 1, 2, .... \tag{S1}$$

where $z = x + iy$ is the complex coordinate in the image plane, $z^* = x - iy$ is its complex conjugate, and the integral is taken along the projected boundary. From the above formula it is evident that all external harmonic moments vanish if the contour is a circle centered at the origin.

For small deviations from a circular shape, $t_n$ is approximately given by the Fourier transform of the contour's polar representation, up to a constant factor. If the enclosed area of the projected *Hydra*'s tissue shape is normalized to π and the contour is described by small distortions of a circle such that its polar representation is given by $r(\theta) = 1 + \delta r(\theta)$ (with $|\delta r(\theta)| \ll 1$), then substituting this representation into Eq. (S1) and expanding to first order in $\delta r(\theta)$ leads to

$$n t_n = \frac{1}{\pi}\int_0^{2\pi} d\theta \delta r(\theta) \exp(-in\theta) = 2\delta r_n^* \tag{S2}$$

Thus, the Fourier coefficients, $\delta r_n$, characterize deformations into *n*-fold symmetric shapes. For instance, $|t_2|$ increases for an elliptical distortion, while $|t_3|$ grows when the contour evolves into a three-petal configuration. The set of harmonic moments, together with the area enclosed by the contour, uniquely define the contour shape. The characterization by harmonic moments is advantageous because it also applies when the polar representation $r(\theta)$ is multivalued and the Fourier coefficients are undefined.

An additional measure of morphology is the shape parameter $\Lambda = 1 - 4\pi A/P^2$, where $A$ is the area and $P$ is the perimeter of the projected tissue [2, 3]. For an ellipse of small eccentricity, $\Lambda \simeq 6t_2^2$. Yet any type of deviation from a circular shape yields a nonzero $\Lambda$.

## Supplementary Note 2: Dimensionality Reduction Procedures

Tracking $t_n$ over time defines the trajectory of the projected tissue image in the morphological space spanned by the harmonic moments (after area normalization). Since the *Hydra*'s body form is approximately a tube-like structure, and early deformations (before the foot is fully formed) occur primarily parallel to the projection plane, harmonic moments effectively capture the evolution of the tissue's three-dimensional structure.

To characterize these trajectories, four dimensionless parameters are used. The first is the duration of the morphological transition— from a near-spherical to an elongated tube-like shape—normalized by the average transition time across all samples. This duration is extracted by fitting $a+b\tanh\left[\left(t-t_*\right)/\sigma\right]$ to $\Lambda(t)$ near the transition region, with $a$ $b$, $t_*$ and $\sigma$ as fitting parameters, and defining $\sigma$ as the approximate transition duration [2]. The remaining three parameters quantify the magnitude of shape fluctuations and are taken as the coefficients of variation (standard deviation over mean) of the absolute values of the first three harmonic moments. These quantify how much the regeneration trajectory meanders in the reduced morphological space spanned by these moments, reflecting the relative time spent in "metastable" states.

Each of the 49 samples in our experiments is thus represented by a four-dimensional vector capturing the main features of its regeneration trajectory. Dimensional reduction is performed first via t-SNE[3] in Mathematica© (perplexity = 12), as shown in Fig. 1, and then via the *Uniform Manifold Approximation and Projection* (UMAP) [4] with Neighbors Number = 14 and Min Distance = 2, illustrated in Fig. S1. Under UMAP, the ratio of the average twin-sample distance to the average non-twin distance is ~0.26, This ratio reduction compared to t-SNE which yields a ratio of ~0.4, results from the more pronounced separation between rapidly transitioning samples and all other samples.

**Supplementary Note 3: Computation of the $Ca^{2+}$ Gradient**

In the first of the two procedures used to identify and compute the gradient in the $Ca^{2+}$ distribution, we take advantage of the foot precursor to determine the polarity axis. We begin by locating the central point of the foot, labeled A in Fig. S8a, and measure the distance of every pixel in the projected image from point A. The pixels used are the ones within the mask polygon, which is contracted by 10% relative to the full tissue's interface to prevent edge effects. These distances are grouped into 50 equal bins, and the fluorescence level in each bin is averaged over 300 consecutive time frames (300 min). The pixel fluorescence intensity is normalized by the maximal pixel intensity in each frame, and the pixel distances are normalized by the scale derived from the tissue's area in each frame. Next, we locate the antipodal point, which is found by drawing a line through point A and the image's center of mass (point B in Fig. S8a) and identifying the point that intersects the image boundary on the opposite side. We repeat the same binning and averaging for this antipodal point. An example of these two calculations is depicted by the red and blue curves in Fig. S8b. Finally, we flip the direction of the distribution associated with point B and then average the two distributions. The resulting black curve in Fig. S8b represents our final estimate of the $Ca^{2+}$ gradient along the line from A to B.

To verify that this procedure indeed captures the gradient along A–B, we repeat the calculation but restrict the averaging only to pixels within a narrow stripe along the A–B line, as shown in Fig. S8c. The stripe computation repeats the same procedure as above, namely, averaging the two distributions within the stripe computed from both ends. The corresponding results, plotted in Fig. S8d, confirm that within the central region of the stripe ( $0.1<x<0.9$ where $x$ is the normalized distance from A along the line to B), the two calculations (red for point A and blue for point B) nearly coincide, and they diverge only near the endpoints. Moreover, the overall results of the two calculations (shown by the black curves in Figs. S8b and S8d) are similar in this central region, as highlighted in Fig. S8e.

The second computational procedure identifies and quantifies the gradient of the average $Ca^{2+}$ density at early times, even before the emergence of the foot precursor, when the projected tissue image is approximately circular and the polarity axis is unknown. We define a sampling circle, with 30 equal-distance vertices, around the image's center of mass, with a diameter equal to 70% of the size determined by the minor axis of the tissue, approximated as an ellipsoid (see Figs. 5a & S8f). Then, along 15 vertices of the sampling circle, each separated by an angle $\Delta\theta = \pi/15$, we measure the fluorescence intensity $\rho(x)$ (as a function of distance $x$ along the diameter), averaged over 50 consecutive time frames (50 min), using the above binning approach for points A and B and averaging of the computed fluorescence distribution from both ends. Here, A and B mark the opposite ends of each diameter (antipodal vertices), and only pixels within the sampling circle are included. A 50-frame average is sufficient to reduce the noise levels and is short enough to ensure that the tissue does not undergo significant shape or rotational changes during the measurement.

Next, we fit $\rho(x)$ to a linear function $\rho_{fit}(x) = \bar{\rho} + \alpha(x - x_0)$, where $\bar{\rho}$ is the average intensity along the diameter, $\alpha$ is a fitting parameter, and $x_0$ is the center point of the diameter. The parameter $\alpha$ represents the average gradient along that diameter. Figure 5b illustrates typical average fluorescence profiles along diameters oriented at different angles. The maximal value of $|\alpha|$ obtained from these fits approximates the $Ca^{2+}$ density gradient (up to a multiplicative constant) along the axis defined by the diameter direction of maximal slope. Accordingly, we define the dimensionless gradient as

$$\text{Dimensionless gradient} = \frac{D}{0.7}\frac{|\alpha|_{\max}}{\bar{\rho}}, \tag{S3}$$

where $D$ is the circle diameter, and $D/0.7$ characterizes the tissue's projected size.

An example of the fluorescence intensity obtained via this procedure is depicted by the blue curve in Fig. S8h. The maximum gradient in that case aligns with the diameter running through the blue cross and the circle's center in Fig. S8f. To verify this method, we repeat the calculation using the stripe geometry depicted in Fig. S8g, producing the brown curve in Fig. S8h. Within the range $0.1 \le x \le 0.9$, the gradients from the two profiles differ by approximately 0.3%, confirming the procedure's validity.

In the second procedure, where the foot precursor is not yet observed and the polarity axis is unknown, several factors can potentially distort the computed $Ca^{2+}$ gradient. First, morphological fluctuations and tissue rotation may lead to an underestimation of the gradient. Although we average over 50 consecutive minutes to minimize these effects, the diameters in the sampling circle remain fixed in orientation. Any residual rotation or shape fluctuation typically lowers the measured gradient; thus, the true gradients are likely larger than our estimates.

A second concern arises from optical projection effects, as discussed in the Supplementary Information of Ref. [4]. When a spherical tissue has a uniform $Ca^{2+}$ activity, its projected fluorescence image often appears as a ring: initially increasing in intensity from the center before fading toward the perimeter. If the sampling circle shifts away from the image center, a spurious fluorescence gradient can result. However, such shifts are generally random and tend to average out over time. A potential exception occurs when the projected tissue shape narrows on one side and widens on the other, shifting the center of mass and producing a false intensity gradient in the absence of a real polarity cue. In practice, these cases usually coincide with the foot precursor's emergence, at which point we can apply the first method (foot-precursor–based) that bypasses this artifact. Moreover, in all tested samples where both methods are applicable, they yield consistent fluorescence profiles and indicate the same polarity direction.

Lastly, tissue orientation within the imaging plane can pose a problem. If the true polarity axis has a significant component perpendicular to the projection plane, even a substantial $Ca^{2+}$ gradient may appear small. However, this situation again leads to an underestimation rather than a false-positive gradient. Our observations indicate that continuous minor shape fluctuations typically maintain the developing body axis parallel to the imaging plane until later stages of regeneration, when the foot has formed, and the axis may rotate out of focus. This loss of focus serves as an indicator that the tissue is rotating toward a perpendicular orientation.

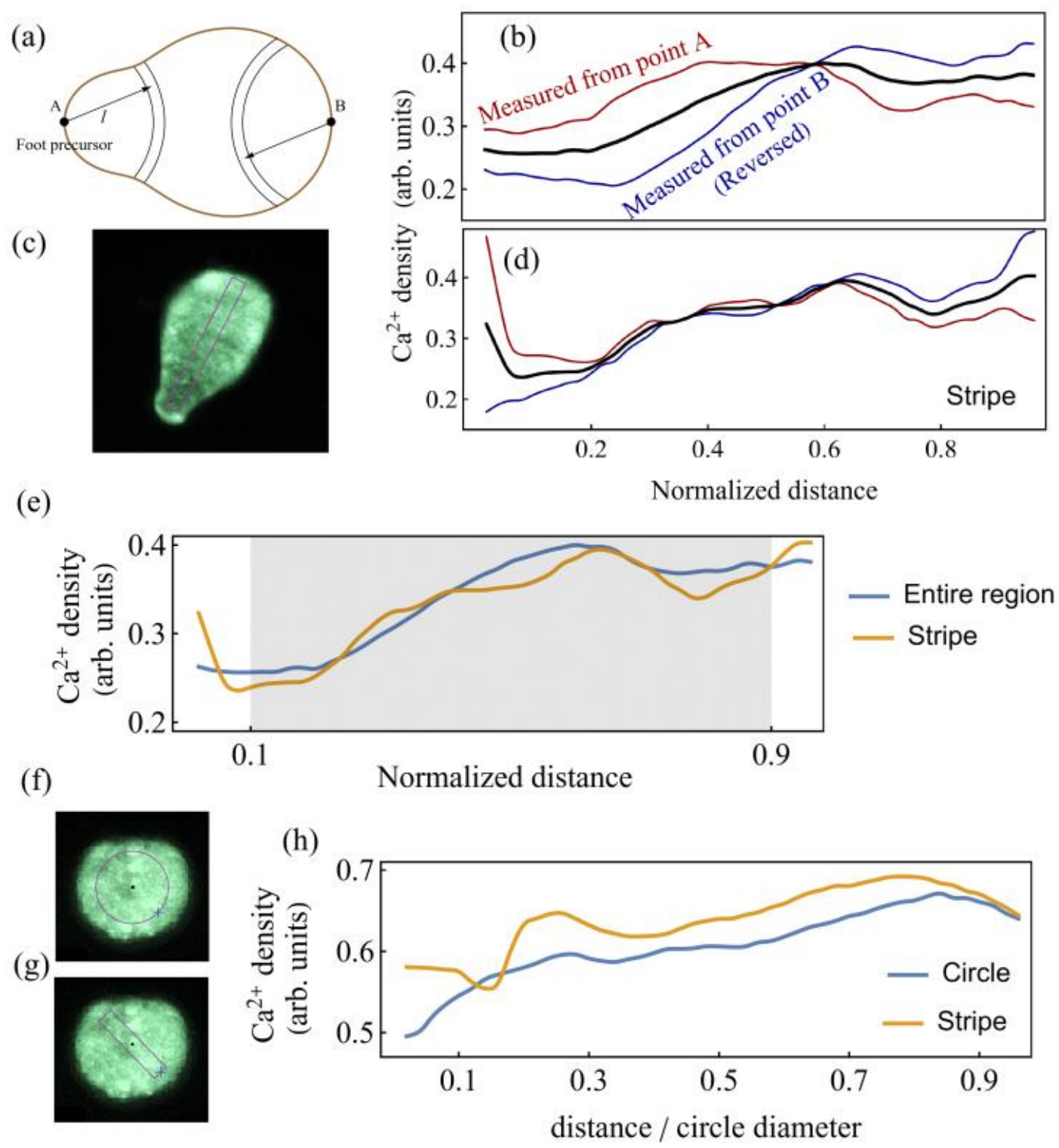

**Fig. S8. Computations of the $Ca^{2+}$ Gradient and Validations.** (a) Schematic illustration of the definition of the foot precursor's center A and its antipodal point B. (b) Example of the measured $Ca^{2+}$ density (arbitrary units) when averaging over the entire projected region of the tissue (within a polygon of its contour reduced by 10%). Red and blue curves show data measured from A and B, respectively, where the distribution computed from B is flipped. The black curve results from averaging the two sets. Here, distance x is the normalized coordinate along the line connecting A and B. (c) Fluorescence image showing the narrow stripe region (outlined in purple) used to verify that the calculation procedure is robust. (d) Measured $Ca^{2+}$ density restricted to this stripe region. Colors match panel (b): red for A, blue for B (flipped), and black for their average. (e) Comparison of the final averaged density profiles from the entire region (blue) and the stripe (brown). The good agreement in the shaded central region, $0.1 < x < 0.9$, confirms that the gradient calculation accurately captures the $Ca^{2+}$ distribution along A–B. (f) and (g) Fluorescence images illustrating the circular (f) and stripe (g) sampling regions used in the second procedure, which is applied when the tissue is approximately circular and the foot precursor has not yet emerged. (h) An example comparing the measured $Ca^{2+}$ density profiles for the circle (blue) and the stripe (brown). The small difference between these profiles validates the consistency of the second procedure for gradient measurements.

## Supplementary Note 4: A Toy Model for Polarity-Morphology Evolution

In this supplementary note, we present a simplified model designed to capture essential dynamical features consistent with experimental observations of *Hydra* regeneration. This toy model extends our previous theoretical framework, which explained the observed morphological transition from an initially spherical tissue fragment to an elongated tubular shape [2].

### Basic Model Description

The previous model applies specifically to regeneration from small tissue fragments taken from the central gastric region of the *Hydra*. It attributes the morphological transition to the interplay between stochastic fluctuations of a scalar field, representing calcium ($Ca^{2+}$) activity, and tissue curvature. The resulting dynamics exhibit a rapid, first-order-like morphological transition.

The dynamics are described by a Langevin-type stochastic equation derived from the effective action [2]:

$$S = S_{el} + S_{\text{ca}} + S_{\text{int}}, \tag{S4}$$

where $S_{el}$ is the elastic free energy:

$$S_{el} = \oiint d^2x\sqrt{g}\kappa\left[H(x) - H_0\right]^2. \tag{S5}$$

Here, $x$ parametrizes the tissue surface, $g$ is the determinant of the metric tensor $g_{\mu\nu}$ describing the tissue geometry, $\kappa$ is the bending modulus, $H$ is the tissue's mean curvature, and $H_0$ is the spontaneous curvature. In passive, thin, homogeneous elastic membranes, bending deformations are typically much less costly than stretching, and neglecting stretching energy is justified only in the effectively un-stretchable limit. This is clearly not the case for *Hydra* tissue, which undergoes substantial stretching during regeneration. This raises the question of why stretching energy is not included in our description. However, one must account for the fact that biological tissues are active materials in which active processes can strongly renormalize the effective elastic moduli. As a result, the effective stretching energy can become considerably smaller than the bending contribution. This issue is discussed in detail in Supplementary Note 5.

The second contribution to the action, $S_{\text{ca}}$, defines the statistical weight of scalar field configurations:

$$S_{\text{ca}} = \oiint d^2x\sqrt{g}\left[\frac{D}{2}g^{\mu\nu}\partial_\mu\phi\partial_\nu\phi + U(\phi)\right], \tag{S6}$$

where the non-negative field $\phi$ represents $Ca^{2+}$ activity, $D$ is a stiffness parameter penalizing large spatial gradients, and $U(\phi)$ is a tilted double-well potential modelling the excitable nature of the tissue. Finally, $S_{\text{int}}$ couples the scalar field to the tissue curvature:

$$S_{\text{int}} = \eta\oiint d^2x\sqrt{g}R^2\phi, \tag{S7}$$

where $R$ is the scalar curvature, and $\eta$ represents the coupling strength.

Upon integrating out the scalar field, the system morphological evolution is described by an effective morphological potential, $V$, via:

$$\exp(-V) = \int \mathcal{D}\phi(x) \exp\left(-\frac{S}{T}\right) \tag{S8}$$

with $T$ acting as an effective temperature associated with stochastic (non-thermal) fluctuations [2]. This effective potential typically displays two minima: one representing the spherical initial state and the other an elongated, tube-like state. A gradual increase in the coupling $\eta$, as suggested by experimental data, eventually shifts the global minimum toward the elongated shape, triggering the observed rapid morphological transition.

In this simplified model, the elongation axis emerges spontaneously through fluctuations. However, experiments indicate that *Hydra* polarity is memorized from the donor animal, necessitating further refinement of the model.

**Extended Model: Inclusion of Polarity Dynamics**

To incorporate inherited polarity, we extend the model by introducing an additional dynamical layer representing positional cues associated with the head and foot. These cues are embodied by two cell types: H-type (head-like) and F-type (foot-like). These cells follow stochastic, cellular-automata-like dynamics, eventually forming clusters at opposite tissue poles and establishing a polarity axis.

In regenerating tissue fragments, these H-type and F-type cells initially lie near each other. During regeneration, they may proliferate, migrate, or be eliminated, ultimately segregating into two distinct clusters. In regeneration from dissociated cells, H- and F-type cells are initially distributed at random.

Communication between these two cell types and the evolving tissue morphology is mediated by two bio-signals, for which H- and F-type cells act as localized sources. H and F cells are assumed to diffuse randomly, biased by the gradients of their signals, and are created and die according to these signals. These bio-signals also modulate the local elastic properties of the tissue and the level of $Ca^{2+}$ activity, and their gradients define the direction of biased diffusion of positional cues. Specifically, they reduce the local bending modulus, increase $Ca^{2+}$ excitability, and generate a gradient in the average $Ca^{2+}$ activity, reflecting the symmetry breaking along the oral–aboral symmetry.

**Qualitative Dynamical Behavior**

In small tissue fragments excised from the gastric region, the numbers of H- and F-type cells are low, and their influence is insufficient to substantially modify the intrinsic activation-like dynamics of the tissue. These dynamics manifest as a rapid transition from a hollow spheroid to an elongated, cigar-like structure. In this case, the primary role of the positional cues is to break spherical symmetry along the polarity axis and thereby set the direction of elongation. Locally reduced bending stiffness in the vicinity of H- and F-type cells promotes the formation of regions of high curvature at the future poles of the cigar-like structure, effectively acting as nucleation sites for axis formation.

For fragments obtained from regions closer to the *Hydra*'s head or foot, abundant H-type or F-type cells alter the local bending modulus and $Ca^{2+}$ activity in a complex manner. Their stronger influence on bending stiffness and $Ca^{2+}$ activity creates a more rugged morphological potential landscape, with much lower barriers, allowing the system to choose trajectories in the morphological space which may transiently occupy intermediate (metastable) morphologies, reflecting a gradual or stepwise progression rather than a sharp transition. Furthermore, differing local $Ca^{2+}$ activities caused by F- or H-type cell distributions generate distinct morphological trajectories depending on the fragment's original location along the axis of the donor *Hydra*.

Finally, in large tissue fragments taken from the central gastric region, the number of positional cues may be somewhat larger than in small fragments, and their contribution still acts primarily as a symmetry-breaking term. However, the main factor underlying the gradual nature of the morphological transition appears to be the increased tissue size, which effectively lowers the morphological barrier below the noise level [5].

**Toy Model Implementation: One-Dimensional Discretized Tissue Loop**

To study these dynamics more explicitly, we model a regenerating *Hydra* fragment as a one-dimensional closed ring of $N$ equally spaced sites with periodic boundary conditions. Small and large tissue fragments correspond to $N = 200$ and $N = 400$ sites, respectively. At each lattice site $i = 1, \cdots, N$, we track (i) a scalar field $\phi_i(t)$, (ii) two diffusive biochemical signals $h_i(t)$ and $f_i(t)$ ("H-signal" and "F-signal"), and (iii) two binary variables $n_{H,i}, n_{F,i} \in \{0,1\}$ indicating the presence of H-type and F-type cells. In addition, the tissue morphology is represented explicitly by a closed curve passing through the $N$ sites at points $R_i(t) = \{X_i(t), Y_i(t)\}$, whose local curvature $H_i(t)$ enters both the mechanical energy and the dynamics of $\phi_i(t)$. The full numerical implementation and parameter values are provided in the Supplementary Software.

The H- and F-signals obey coupled reaction–diffusion dynamics on the ring,

$$\begin{aligned} \frac{dh_i}{dt} &= D_H \Delta h_i + \rho_H n_{H,i} - \kappa_{ann} h_i f_i - \gamma_H h_i \\ \frac{df_i}{dt} &= D_F \Delta f_i + \rho_F n_{F,i} - \kappa_{ann} h_i f_i - \gamma_F f_i \end{aligned} \tag{S9}$$

where $D_H \gg D_F$ are the diffusivities of the H- and F-signals, $\rho_H$ and $\rho_F$ are production rates by H- and F-cells, $\gamma_H$ and $\gamma_F$ are linear decay rates, and $\kappa_{ann}$ controls mutual annihilation of the two signals. The discrete Laplacian $\Delta$ is implemented spectrally with periodic boundary conditions.

The dynamics of $\phi_i$ is modeled as a noisy gradient flow in an effective tilted double-well potential. In the absence of signals and curvature, this potential has two minima at $\phi = \bar{\phi} \pm \Delta\phi/2$, where $\bar{\phi}$ is the midpoint and $\Delta\phi$ is the separation between low- and high-activation states. The evolution equation is

$$\frac{d\phi_i}{dt} = -\frac{\partial U}{\partial \phi_i} + \sqrt{2T_0}\,\xi(t), \tag{S10}$$

with

$$U(\phi_i) = \frac{\alpha(f_i, h_i)}{4}\left[(\phi_i - \bar{\phi})^2 - \frac{\Delta\phi^2}{4}\right]^2 + \left(\lambda_0 + \lambda_1 f_i + \eta H_i^2\right)\phi_i \tag{S11}$$

Here $\alpha(f,h) = \alpha_0 - 2\sigma\left[(h+f)/w_\phi\right]$ is a signal-dependent quartic prefactor defined through a smooth sigmoidal function $\sigma(z) = \left[1 + \tanh(z)\right]/2$, $\lambda_0$ is a baseline tilt, $\lambda_1 f$ is an F-signal–dependent ("foot biosignal") tilt, $\eta(t)H_i^2$ couples $\phi_i$ to the local squared curvature $H_i^2$, and $\xi_i(t)$ is a Gaussian white noise with zero mean and unit variance, with $T_0$ setting the noise amplitude. In the numerical implementation, we additionally impose the constraint $\phi_i(t) \geq 0$ at all sites and times, reflecting that $\phi$ represents a non-negative $Ca^{2+}$ activity level.

We deliberately omit explicit diffusion of $\phi$ between neighboring sites (no Laplacian of $\phi$), because $\phi_i$ represents local $Ca^{2+}$ activity. Direct intercellular diffusion of free $Ca^{2+}$ through gap junctions is strongly limited by buffering and active pumping; intercellular $Ca^{2+}$ expansion are thought to propagate mainly via diffusing upstream messengers (e.g. $IP_3$) or ATP release, rather than by substantial $Ca^{2+}$ flux itself [6, 7, 8]. Thus, spatial coupling of $\phi$ in the model arises indirectly via the reaction–diffusion dynamics of $h, f$ and the curvature-dependent mechanical feedback, rather than through direct diffusion of the $Ca^{2+}$ activity field.

The linear $\phi$ term in Eq. (7) sets the tilt of the double-well. When the foot biosignal $f_i$ is large, or when the local curvature $H_i^2$ is large, this tilt disfavors occupation of the higher-activation minimum and thereby inhibits excitation into the high-$\phi$ state. This implements two independent control channels for $Ca^{2+}$ activity: a chemical channel, mediated by the F-signal, and a mechanical channel, mediated by curvature.

In addition, the dependence of the quartic prefactor $\alpha(h, f)$ on the combined biosignals is chosen such that $\alpha$ decreases in regions where $h + f$ is large, thereby lowering the local potential barrier between the two minima. This makes transitions between low and high $Ca^{2+}$ activity more likely in signal-rich regions and, through the curvature-$\phi$ coupling in the mechanical energy, modifies the local mechanical response of the tissue by facilitating shape changes precisely where the biochemical cues are strong.

H- and F-cells perform biased random walks on the ring and undergo signal-dependent birth and death. H-cells are biased up the gradient of $h$, whereas F-cells are biased down the gradient of $h$. At each step, a randomly chosen cell attempts a hop over a distance drawn from a narrow exponential distribution; the hop direction is chosen probabilistically using a tanh-nonlinearity of the local difference in $h$ between left and right neighbors (up or down $\nabla h$). Moves are executed only if the target site is empty of both cell types.

Birth and death probabilities are smooth functions of the local signal levels. H-cells are created above a threshold in $h$ and removed when $h$ leaves a prescribed fuzzy band $[H_{low}, H_{high}]$; F-cells obey analogous rules controlled by $f$ and a band $[f_{low}, f_{high}]$. These rules enforce mutual exclusion of H- and F-cells at each site and implement soft, signal-dependent maintenance of H- and F-type domains. The explicit sigmoidal rate functions and parameter values are given in the Supplementary Software.

The 1D tissue loop is represented by $N$ points $R_i(t) = \{X_i(t), Y_i(t)\}$ on a closed curve of fixed area (initially close to a circle). Local curvature $H_i$ is computed from triplets of neighboring points, with regularization to avoid numerical singularities. The mechanical energy is

$$F = \sum_{i=1}^{N} \left[ \kappa_{eff,i} (H_i - H_M)^2 + \eta(t) H_i^2 \phi_i \right] \Delta s, \tag{S12}$$

where $H_M$ is a preferred curvature, $\Delta s$ is the local arc-length step, and

$$\kappa_{eff,i} = \kappa \exp\left(-\frac{h_i}{\kappa_{sens}}\right) \tag{S13}$$

is an effective bending rigidity that is reduced in regions of high H-signal, making head-like regions mechanically softer, in accordance with the experimental observation of Ref. [9]. Here $\eta(t)$ controls the strength of the curvature–activation coupling. In the simulations we take

$$\eta(t) = \eta_{initial} + \Delta\eta\sigma\left(\frac{2t}{n_{frames}}\right) \tag{S14}$$

$n_{frames}$ is the number of time frames in the simulation (which is identical for all initial conditions). Both the initial value $\eta_{initial}$ and the ramp $\Delta\eta$, can differ between simulation protocols, as detailed below. The time dependence of the coupling parameter in Eq. (S14) is used to represent "hidden" internal dynamics that gradually strengthen the curvature–activation coupling, for example, due to improved alignment of supra-cellular actin fibers or more efficient operation of myosin motors. In our simulations, this gradual improvement is assumed to be more pronounced for tissue fragments taken from the central gastric region than for fragments taken near the head or near the foot.

Shape updates are proposed by correlated normal displacements of the boundary (via a smooth kernel on the ring) followed by recentering, rescaling to restore unit area, and equidistant re-parametrization. Each proposed shape is accepted or rejected using a Metropolis criterion at an effective Monte Carlo temperature $T_{MC}$, with the proposal amplitude adaptively tuned to maintain a target acceptance rate. Algorithmic details are provided in the Supplementary Software.

The simulations use a two-stage initialization. First, we construct a "bare" configuration; second, we let the biochemical and activation fields relax for a finite transient while keeping the geometry and cell masks fixed. The state at the end of this transient is taken as the initial condition for the main dynamics.

In the first stage, we start from a nearly circular ring of unit area. The curvature profile of this reference shape is computed once and kept fixed during the pre-relaxation stage. The activation field $\phi_i$ is initialized close to the lower minimum of the double-well and shifted by small Gaussian fluctuations. The signals $h_i$ and $f_i$ are initialized as low, nearly homogeneous fields with weak noise (small positive values plus Gaussian fluctuations, clipped at zero). H- and F-type cells are placed on the ring in approximately disjoint bands using binary masks, with at least one cell of each type enforced. Different biological scenarios are encoded by choosing different band widths, shifts, and occupancy probabilities:

- Head-side fragments: H-cells occupy a relatively wide arc with high occupancy probability, while F-cells occupy a narrower band shifted along the ring, yielding an H-dominated domain and a weaker F-domain.
- Foot-side fragments: F-cells form the broader, more densely occupied domain, while H-cells occupy a narrower band, inverting the head-like pattern.
- Central fragments (small and large): H- and F-cells occupy bands that are approximately symmetric around the "central" region of the ring, with comparable occupancy fractions. "Small center" and “large center” cases differ by system size ( $N = 200$ vs $N = 400$ ) and slightly different occupancy probabilities, corresponding to fragments cut from narrower vs broader regions of the gastric column.

For each of these cases, we also assign a specific initial value of the curvature–activation coupling, $\eta_{initial}$, and (where relevant) a ramp amplitude $\Delta\eta$. Thus, both the H/F-cell pattern and the initial mechanical sensitivity to curvature are part of the prescribed initial condition.

Before starting the main loop, we evolve $h_i$, $f_i$, and $\phi_i$ for a fixed number of time steps (5000 iterations in the code) while keeping the tissue geometry, curvature, and the initial distribution of the F- and H-type cells fixed to the initial ring. During this stage, the initially almost homogeneous $h$, $f$, and $\phi$ fields relax into non-trivial profiles shaped by the imposed H/F-cell arrangement and the chosen $\eta_{initial}$, while the tissue shape remains nearly circular. The configuration of the aforementioned field, at the end of this pre-relaxation period, serves as the true initial condition for the subsequent coupled evolution of geometry, fields, and cells in the main simulation.

**Results**

For each of the four classes of initial conditions described above, we performed simulations for seven slightly different realizations (differing by noise and microscopic initialization). The results for initial conditions designed to mimic small tissue fragments from the central gastric region are summarized in Fig. S9.

Figure S9a shows a typical time trace of the shape parameter, exhibiting an abrupt morphological transition from a nearly circular state to an elongated, cigar-like shape, in close analogy to the experimental observations (Figs. 2a and S2). The additional six traces from different realizations shown in Fig. S9e, display similar rapid transitions, while at the same time illustrating the substantial variability in the precise timing and details of the transition.

Figure S9b presents snapshots of the tissue morphology together with the evolving positional cues, represented by red and blue bars corresponding to the F- and H-type cell populations, respectively. Over time, these cues segregate and accumulate at two opposite sides of the tissue, thereby defining a polarity axis that, in turn, specifies the emergent morphological axis.

Figure S9c shows the time-averaged intensity of the $\phi$-field along the polarity axis, averaged over the first 200 time frames. A clear gradient is observed, closely resembling the experimental profile (Fig. 4d). Figure S9d displays the $\phi$ -field intensity at the final time point, which qualitatively resembles the endpoint patterns seen in the insets of Fig. 4d.

Figures S10, S11, and S12 show analogous results for simulations initialized to mimic (i) large tissue fragments from the central gastric region, (ii) small near-foot fragments, and (iii) small near-head fragments, respectively. In these cases, the shape parameter exhibits a gradual morphological transition, in agreement with the experimental data in Figs. 2c, 3c, and 3d. Taken together, these comparisons demonstrate that our toy model captures the main qualitative features of the morphological transitions and supports the view that the nature of the transition depends on both tissue size and positional cues inherited from the parent *Hydra*.

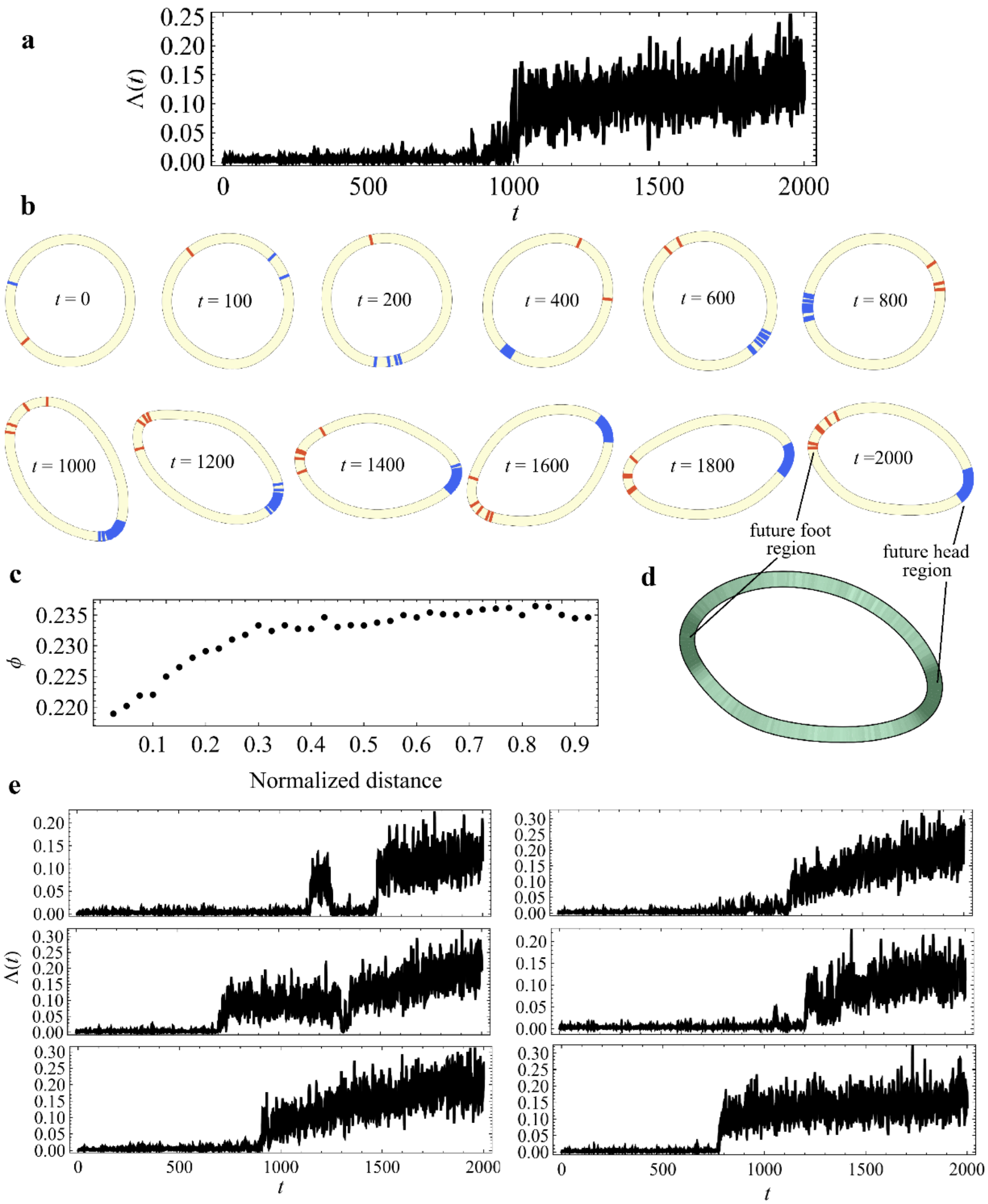

**Fig. S9. Toy-model dynamics for small tissue fragments from the central gastric region.** (a) Time evolution of the shape parameter for a representative simulation, showing an abrupt morphological transition from a nearly circular state to an elongated, cigar-like shape, analogous to the behavior observed experimentally.(b) Snapshots of the tissue morphology at successive times, together with the evolving positional cues: red and blue bars indicate F-type and H-type cell bands, respectively. Over time these cues segregate to opposite sides of the tissue, thereby defining the polarity axis and, consequently, the morphological axis.(c) Time-averaged of $\phi$ field ($Ca^{2+}$ activity) along the polarity axis, averaged over the first 200 time frames, showing a clear gradient along the axis.(d) The $\phi$ field profile at the final simulation time, illustrating the endpoint pattern of excitation. (e) Time traces of the shape parameter for six additional simulations with slightly different realizations of the initial conditions and noise, demonstrating that the transition is consistently rapid while its timing and detailed trajectory are variable.

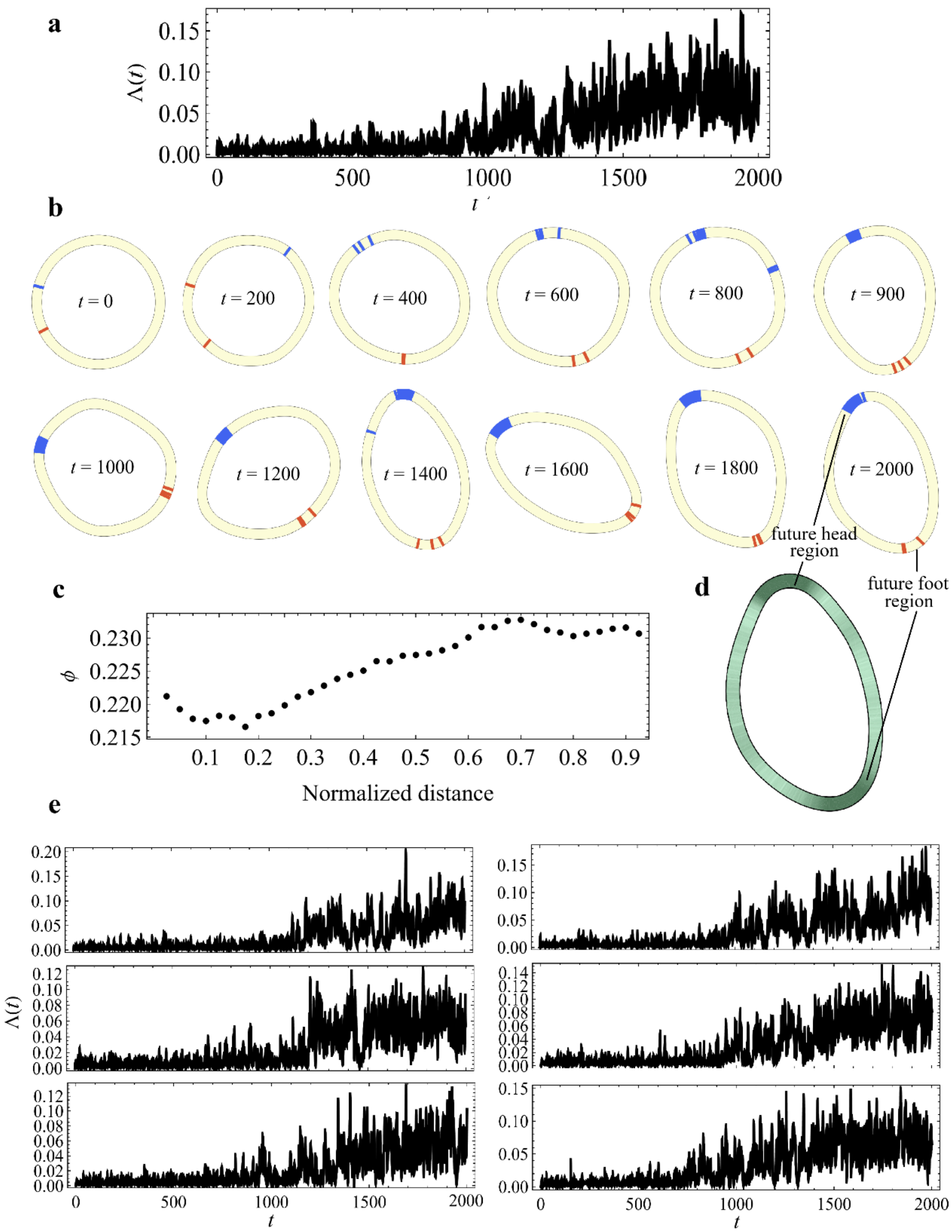

**Fig. S10. Toy-model dynamics for large tissue fragments from the central gastric region.** Same as Fig. S9, but for initial conditions that represent large tissue fragments taken from the central gastric region.

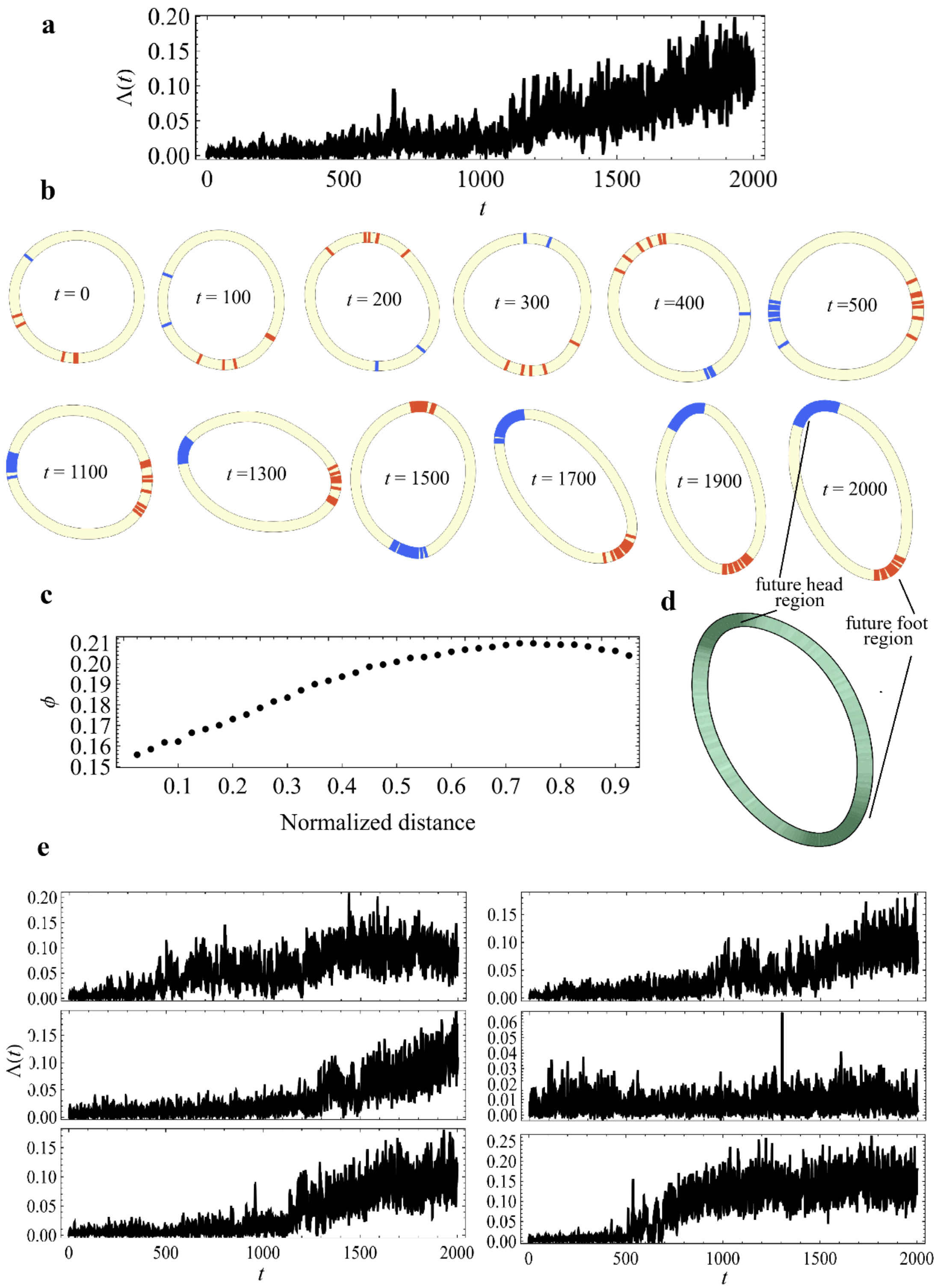

**Fig. S11. Toy-model dynamics for small near-foot tissue fragments.** Same as Fig. S9, but for initial conditions that represent small near-foot tissue fragments.

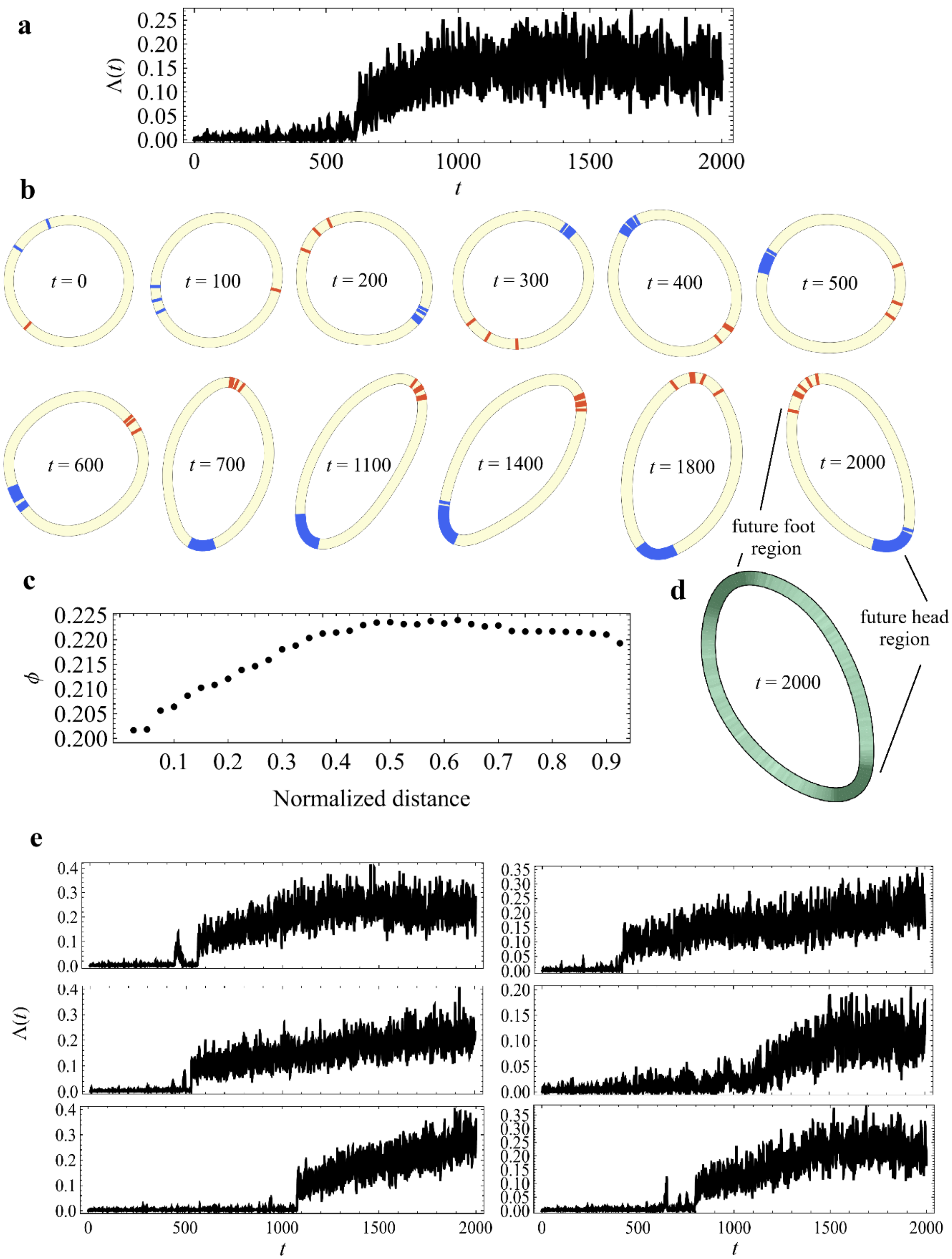

**Fig. S12. Toy-model dynamics for small near-head tissue fragments.** Same as Fig. S9, for but for initial conditions that represent small near-head tissue fragments.

**Supplementary Note 5: Mechanical Asymmetry in Stretching and Bending of *Hydra* Tissue**

*Hydra* tissue is organized as a bilayer of epithelial sheets, an outer ectoderm and an inner endoderm, separated by a thin extracellular matrix known as the mesoglea. The two layers differ in their cellular geometry: ectodermal cells are approximately isotropic, with an aspect ratio near 1, whereas endodermal cells are elongated along the apico-basal axis with an aspect ratio of 4–5. Despite this structural heterogeneity, the two epithelial layers remain mechanically coupled and deform together during morphogenesis. For clarity and tractability in what follows, we approximate this bilayer as an effective monolayer with coarse-grained mechanical properties [10].

A striking feature of *Hydra* morphogenesis is the tissue's ability to undergo rapid and large-scale shape changes. During regeneration, for example, the body column can elongate dramatically within a timescale of minutes. Similarly, during feeding or spontaneous behavior, *Hydra* can extend its tentacles and body column in seconds. The electrically excitable nature of the *Hydra* epithelium that behaves as a muscle enables these rapid morphological changes. These processes in turn, involve a substantial increase in surface area, yet proceed without apparent membrane rupture or visible signs of membrane biosynthesis. These observations suggest that *Hydra* tissue contains a mechanism for rapid membrane deployment, allowing the cells to expand their basal and apical surfaces on demand.

In many epithelial systems, mechanical flexibility is provided by membrane reservoirs. On the apical side, microvilli (actin-supported protrusions) store excess membrane that can flatten under increased tension and release area without de novo biosynthesis [11]. Deployment can occur without immediate actin depolymerization; core actin bundles may persist or remodel later, enabling rapid, energetically modest expansion of surface area [12, 13]. Additional apical reservoirs include sub-apical folds and caveolae, which are recruited under mechanical load and act as tension buffers [14]. The basal surface, on the other hand, typically exhibits membrane infoldings (the *basal labyrinth*) rather than microvilli, providing additional area buffering at the cell–matrix interface [14].

While direct experimental evidence for membrane reservoirs in *Hydra* epithelial cells is currently lacking, the mechanical behavior of *Hydra* tissue strongly suggests their functional presence. The speed and extent of tissue deformation (on the order of seconds to minutes and involving doubling or more of surface area) argue for a pre-existing, deployable membrane reserve. Whether this is mediated by microvilli, folds, or other structures remains to be determined experimentally [12, 13].

To explore the mechanical consequences of the membrane reservoir, we introduce a minimal one-dimensional model in which the tissue is represented as a chain of elastic elements, each corresponding to a segment of the epithelial sheet (Fig. S13a). In the undeformed state, each element contains membrane slack, illustrated schematically as a wavy contour (Fig. S13b). When the tissue is stretched (Fig. S13d), this slack is straightened out, allowing the segment to elongate with minimal energetic cost. In contrast, compressing the element (Fig. S13f) requires further folding or crowding of already-compacted membrane and cytoskeletal components, which is energetically unfavourable due to steric constraints and cytoskeletal stiffness.

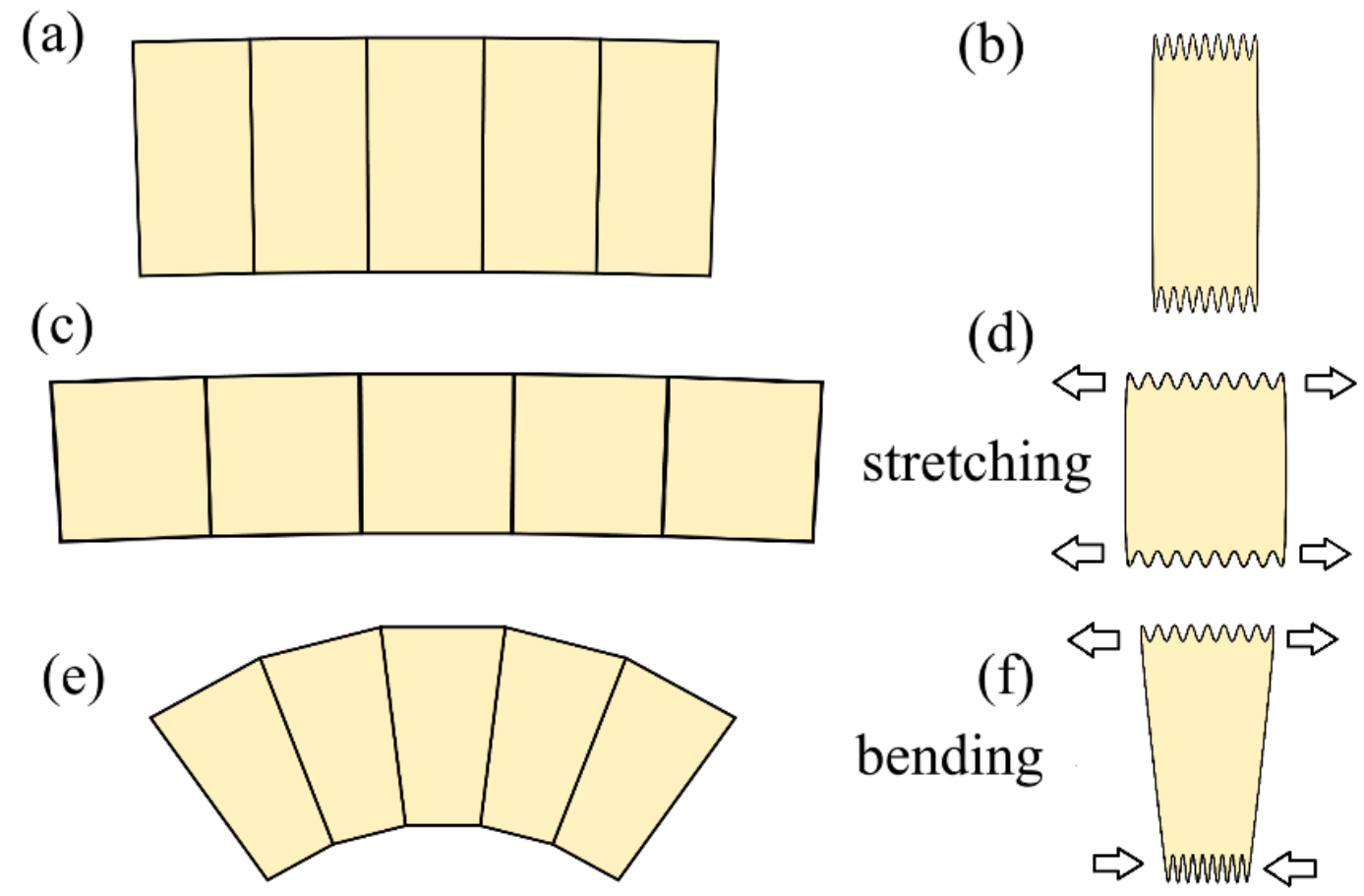

**Figure S13. Schematic of asymmetric mechanics in a slack-bearing epithelium.** (a) Effective 1D tiling of the tissue in the undeformed state. (b) A single segment in cross-section, with apical and basal membrane undulations (a "reservoir" of excess area) represented as wavy contours. (c) Axial stretching at the tissue level: segments elongate while keeping thickness nearly unchanged. (d) Under stretch, the undulations unfold, providing additional membrane with little energetic cost (arrows indicate applied tensile load). (e) Bending at the tissue level: outer segments extend while inner segments compress. (f) Under compression, undulations must be compacted and new folds created, which is energetically costly (arrows indicate compressive load). The contrast between facile unfolding under tension and costly compaction under compression produces a strong mechanical asymmetry: stretching is soft, whereas bending is hard.

This leads to a mechanical asymmetry: the tissue resists compression but shows little resistance to extension. Stretching deformations (Fig. S13c) are energetically inexpensive, as they can be accommodated passively by deploying membrane slack. In contrast, bending deformations (Fig. S13e), which involve extension of the outer arc and compression of the inner arc, result in a significant energy cost dominated by the compressive side. As we explain in detail, this asymmetry justifies neglecting the stretching term in the elastic energy of the tissue and retaining only a Helfrich-like bending term (Eq. S5), which captures the dominant energetic contribution associated with morphological changes that occur during regeneration.

To model this asymmetry quantitatively, we adopt a phenomenological description for the fraction of membrane slack recruited, $f(\varepsilon)$, as a function of the local strain $\varepsilon$. Each coarse-grained segment is assumed to contain many membrane undulations that can be progressively deployed under tension and compacted under compression. A convenient and analytically tractable choice that captures these features is an exponential recruitment law characterized by a strain scale $\varepsilon_c$:

$$f(\varepsilon)=1-\exp\left(-\frac{\varepsilon}{\varepsilon_c}\right) . \tag{S15}$$

This choice is motivated by the following complementary considerations:

(i) Maximum-entropy argument [15]: Membrane reservoirs, such as folds or curves, are typically stabilized by energy barriers arising from protein scaffolds, and membrane–cortex adhesions. Their deployment under increased tension is therefore plausibly an activation-over-a-barrier process [16, 17, 18]. If local deployment thresholds $\theta > 0$ are only constrained by their mean $\langle\theta\rangle = \varepsilon_c$, maximizing entropy over the distribution $p(\theta)$ yields $p(\theta)=\varepsilon_c^{-1}\exp(-\varepsilon/\varepsilon_c)$. The resulting cumulative recruitment by strain $\varepsilon$ is $f(\varepsilon)=\Pr[\theta<\varepsilon]=1-\exp(-\varepsilon/\varepsilon_c)$.

(ii) Stochastic recruitment with a constant opening rate [19]: Let us model the tissue as containing many independent slack-bearing units, each of which can open at most once during the deformation. Over the time/strain window of interest we assume: (a) once a unit opens it does not close again (or re-closing is negligible), (b) different units behave independently, and (c) each closed unit has the same fixed probability per unit strain to open, regardless of how much strain it has already experienced. In other words, during an infinitesimal strain increment $d\varepsilon$, a closed unit opens with probability $hd\varepsilon$, independent of its prior history. Under these assumptions, the survival of a single unit (still closed at strain $\varepsilon$) is

$$S(\varepsilon)=\exp\left(-\int_0^{\varepsilon} hd\varepsilon'\right)=\exp(-h\varepsilon) \tag{S16}$$

Therefore, the probability that the reservoir is opened $f(\varepsilon)=1-S(\varepsilon)=1-\exp(-\varepsilon/\varepsilon_c)$, where we set $h=\varepsilon_c^{-1}$. This constant-hazard picture is equivalent to assuming exponentially distributed thresholds as described above.

(iii) Extreme-value coarse-graining [20]: Assume each coarse-grained patch contains $m$ independent subdomains, such that the $j$-th subdomain has a local opening threshold $\theta_j$ drawn from a common parent distribution with cumulative distribution function (CDF) $F(\varepsilon)=\Pr(\theta<\varepsilon)$. The patch is assumed to be deployed once the first subdomain opens, i.e., when $\varepsilon$ exceeds $\theta_{\min}=\min\theta_j$. If $F(\varepsilon)$ is regular near $\varepsilon\to 0$, i.e. $F(\varepsilon)\approx c\varepsilon$, where $c$ is a constant, the minimum threshold CDF yields an approximate exponential behavior, $\Pr[\theta_{\min}<\varepsilon]=1-[1-F(\varepsilon)]^m\approx 1-\exp(-mc\varepsilon)$, with $\varepsilon_c^{-1}=mc$.

Multiplying $f(\varepsilon)$ by the total amount of available slack in the unstrained state, $N_0$, yields the expected slack deployed strain $\varepsilon$:

$$N(\varepsilon)=N_0\left[1-\exp\left(-\frac{\varepsilon}{\varepsilon_c}\right)\right]. \tag{S17}$$

This function increases approximately linearly for small strain $|\varepsilon|\ll\varepsilon_c$, and saturates at large positive strain $\varepsilon\gg\varepsilon_c$, consistent with full deployment of the available slack. For negative strain, $\varepsilon<0$, the function, $N(\varepsilon)$ becomes increasingly negative. This case should be interpreted phenomenologically as the additional membrane slack that must be created/inserted to accommodate the same area within a reduced spatial extent.

The elastic energy associated with either the basal or apical side of the segment is, then, given by:

$$E(\varepsilon)=\frac{1}{2}kN^2(\varepsilon) \tag{S18}$$

where $k$ is an effective stiffness coefficient. This functional form represents the simplest quadratic dependence of energy on deviation from equilibrium. The energy is minimized when no slack is recruited relative to the unstrained state, i.e., when $N(\varepsilon)=0$ (which occurs at $\varepsilon=0$). Deviations from this state, either by stretching ($N(\varepsilon)>0$) or compression ($N(\varepsilon)<0$), require energy. However, due to the asymmetry in the form of $N(\varepsilon)$, the energy cost is highly asymmetric when $|\varepsilon|>\varepsilon_c$: stretching deformations lead to a gradual, saturating increase in energy, whereas compressive deformations result in a steep and rapidly growing energy penalty.

For stretching, we assume that the outer and inner faces of the tissue segment experience the same positive strain, $\varepsilon$, thus $E_{stretch}=2E(\varepsilon)$. For large strain (which is the typical situation during the morphological transition in *Hydra* regeneration):

$$E_{stretch} \xrightarrow[\varepsilon \gg \varepsilon_c]{} kN_0^2 \ . \tag{S19}$$

The constant offset $kN_0^2$ at large tension is physically irrelevant and can be subtracted; what matters is that changes in the elastic energy, $\Delta E_{stretch}$, due to the tissue stretching (when deformation is large), become negligibly small.

For bending, a segment of thickness $w$ experiences tensile and compressive strains at the outer and inner faces,

$$\varepsilon_{\pm} = \pm wH/2 , \tag{S20}$$

with $H$ the local curvature. Then, the total bending energy of the segment follows from the sum of the face contributions,

$$E_{bend}(H) = \frac{k}{2}N^2(\varepsilon_+) + \frac{k}{2}N^2(\varepsilon_-) \tag{S21}$$

At large curvature $wH/\varepsilon_c \gg 1$, the tensile side saturates ($N(\varepsilon_+) \to N_0$), while the compressive side dominates (as it grows exponentially with the strain) yielding the asymptotic form

$$E_{bend}(H) \approx \frac{k}{2}N_0^2 \exp\left(\frac{wH}{\varepsilon_c}\right). \tag{S22}$$

This behavior supports treating stretching as soft (saturating) and bending as the dominant elastic cost. In an intact regenerating *Hydra*, global constraints- closure of the epithelial shell, near-inextensibility of the surface, and hydrostatic pressure of the enclosed fluid - impose a relationship between curvature and load [13, 14]. We treat this by minimizing the constrained free energy

$$\tilde{E}(H,\lambda) = E_{bend}(H) - \lambda\left(H - \bar{H}\right), \tag{S23}$$

where $\lambda$ is a Lagrange multiplier representing the external bending moment needed to enforce the global constraint, and $\bar{H}$ is the corresponding target/average curvature set by the global constraint. Let $H_*$ denote the stationary curvature of the constrained problem, i.e. $d\tilde{E}/dH\big|_{H=H_*}=0$. A local expansion about $H_*$ removes the linear term by construction and yields a Helfrich-like quadratic response [21],

$$\tilde{E}(H) \simeq \tilde{E}_0 + \frac{1}{2}\kappa_{eff}(H - H_*)^2, \quad \text{with} \quad \kappa_{eff} = \left.\frac{d^2 E_{bend}}{dH^2}\right|_{H=H_*}. \tag{S24}$$

Using the large-curvature asymptotic result (S22) gives a large effective bending modulus,

$$\kappa_{eff} = \frac{k}{2}\left(\frac{N_0 w}{\varepsilon_c}\right)^2 \exp\left(\frac{wH_*}{\varepsilon_c}\right). \tag{S25}$$

reflecting the steep energetic cost of compression in the inner arc. Thus, under realistic global constraints, the tissue exhibits a Helfrich-like quadratic response around $H_*$ with a large prefactor, while pure stretching remains soft due to slack deployment.

Taken together, these results support the continuum approximation in which stretching contributions are neglected and the tissue's elasticity is described by a Helfrich-like bending term. They also imply that biological signals producing only modest changes in the characteristic strain scale $\varepsilon_c$ can efficiently tune the bending modulus because $\kappa_{eff}$ is exponentially sensitive to $\varepsilon_c$ (see Eq. S25). This sensitivity is consistent with the assumption in Supplementary Note 3 that signals carried by units with positional cues affect the bending modulus, thereby influencing polarity–morphology evolution.